\documentclass[aps,prx,showpacs,twocolumn,superscriptaddress,longbibliography]{revtex4-2}

\usepackage[english]{babel}

\usepackage{graphicx}
\usepackage{dcolumn}
\usepackage{bm}

\usepackage{longtable}
\usepackage{array}
\usepackage{booktabs} 
\usepackage{fontawesome} 
\usepackage{amsmath}
\usepackage{fdsymbol}
\usepackage[normalem]{ulem}

\def\aap{Astron. Astrophys.} 
\def\araa{ARA\&A}  
\def\procspie{Proc.~SPIE}   
\def\mnras{Mon. Not. R. Astron. Soc.}           
\def\pasp{Publ. Astron. Soc. Pac.}               
\def\aj{Astron. J.}               
\def\pasj{Publ. Astron. Soc. Jpn.}

\newcommand{\nn}{$^{14}$N$^{14}$N}
\newcommand{\nnp}{$^{14}$N$^{15}$N}
\newcommand{\oo}{$^{16}$O$^{16}$O}
\newcommand{\oop}{$^{16}$O$^{17}$O}
\newcommand{\oopp}{$^{16}$O$^{18}$O}
\newcommand{\coo}{$^{12}$C$^{16}$O$^{16}$O}

\newcommand{\espr}{ESPRESSO}
\newcommand{\muse}{MUSE}
\newcommand{\harmoni}{HARMONI}
\newcommand{\fors}{FORS2}

\newcommand\T{\rule{0pt}{2.6ex}}       
\newcommand\B{\rule[-1.2ex]{0pt}{0pt}} 

\newcommand{\lbeam}{\ast}

\begin{document}


\title{The pure-rotational and rotational-vibrational Raman spectrum of the atmosphere at an altitude of 23\,km}

\author{Fr\'ed\'eric\,P.\,A.\,Vogt}
\email{frederic.vogt@alumni.anu.edu.au}
\homepage{https://fpavogt.github.io}
\affiliation{Federal Office of Meteorology and Climatology MeteoSwiss, Chemin de l'A\'erologie 1, 1530 Payerne, Switzerland.}

\author{Andrea Mehner}
\affiliation{European Southern Observatory (ESO), Av. Alonso de C\'ordova 3107, 763 0355 Vitacura, Santiago, Chile.}
\author{Pedro Figueira}
\affiliation{Department of Astronomy, University of Geneva, Chemin Pegasi 51, Versoix, Switzerland.}
\affiliation{Instituto de Astrofíısica e Ciências do Espaço, Universidade do Porto, CAUP, Rua das Estrelas, 4150-762 Porto, Portugal.}

\author{Shanshan Yu}
\affiliation{Jet Propulsion Laboratory, California Institute of Technology, Pasadena, CA, 91109, USA.}

\author{Florian Kerber}
\author{Thomas Pfrommer}
\author{Wolfgang Hackenberg}
\author{Domenico Bonaccini Calia}
\affiliation{European Southern Observatory (ESO), Karl-Schwarzschild-Str. 2, 85748 Garching, Germany}

\date{\today}

\begin{abstract}
Ground-based optical astronomical observations supported by or in the vicinity of laser guide-star systems can be contaminated by Raman-scattered laser photons. Anticipating, alleviating, and correcting for the impact of this self-inflicted contamination requires a detailed knowledge of the pure-rotational and rotational-vibrational spectrum of the molecules in our atmosphere. We present a 15.3\,hr-deep combined spectrum of the 4LGSF's 589\,nm$\approx$509\,THz sodium laser beams of Paranal observatory, acquired with the {\espr} spectrograph at a resolution $\lambda/\Delta\lambda\cong140'000\approx0.12$\,cm$^{-1}$ and an altitude of 23\,km above mean sea level. We identify 865 Raman lines over the spectral range of [3770; 7900]\,\AA$\approx$[+9540; -4315]\,cm$^{-1}$, with relative intensities spanning $\sim$5 orders of magnitudes. These lines are associated to the most abundant molecules of dry air, including their isotopes: \nn, \nnp, \oo, \oop, \oopp, and \coo. The signal-to-noise of these observations implies that professional observatories can treat the resulting catalogue of Raman lines as exhaustive (for the detected molecules, over the observed Raman shift range) for the purpose of predicting/correcting/exploiting Raman lines in astronomical data.\\

Our observations also reveal that the four laser units of the 4LGSF do not all lase at the same central wavelength. We measure a blueshift of $+43\pm10$\,MHz $\cong-50\mp10$\,fm with respect to $\lambda_{\lbeam}=5891.59120$\,\AA\ for LGSU1/2, and $+94\pm10$\,MHz $\cong-109 \mp 10$\,fm for LGSU3/4. These offsets, including the difference of $\sim$50\,MHz between LGSU1/2 and LGSU3/4, are larger than the observed 4LGSF spectral stability of $\pm3$\,MHz over hours. They remain well within the operational requirements for creating artificial laser guide-stars, but hinder the assessment of the radial velocity accuracy of {\espr} at the required level of 10\,m\,s$^{-1}$. Altogether, our observations demonstrate how Raman lines can be exploited by professional observatories as highly-accurate, on-sky wavelength references.\\

\noindent{\bf Popular summary\\}

Professional ground-based astronomical telescopes and instruments can be equipped with special optical systems designed to alleviate the impact of atmospheric turbulence on the observations. A growing number of these systems rely on 589\,nm$\approx$509\,THz lasers to excite sodium atoms in the mesosphere (at an altitude of $\sim$80-100\,km), causing them to glow. This allows astronomers to create ``artificial laser guide-stars'' at any location in the sky, that can be used to characterize atmospheric turbulence. As these laser photons propagate through the lower, denser, layers of the atmosphere while on their way up to the mesosphere, they may interact with air molecules and dust particles in different ways, including via Raman scattering which is a so-called ``inelastic'' mechanism. Raman-scattered laser photons loose (resp. gain) energy by (de-)exciting pure-rotational and/or rotational-vibrational transitions from air molecules. \\

The direct consequence of Raman scattering is that the up-link beams from laser guide-star systems have a very complex spectral signature, comprised of numerous emission lines. In this article, we present the most accurate observations of these Raman emission lines acquired with an astronomical instrument to-date. The resulting catalogue can be used by astronomers to correct their contaminated observations using post-processing software. Our observations also reveal that the four laser systems that are installed at the observatory of Cerro Paranal in Chile do not all have the exact same wavelength. Although this discovery limits our ability to independently validate the spectral accuracy of the most stable astronomical spectrograph at Cerro Paranal, it also unambiguously demonstrates the unique potential of exploiting Raman lines as accurate wavelength references at professional astronomical observatories.
\end{abstract}

\maketitle


\section{Introduction}

Laser guide-star systems, supporting adaptative optical techniques, allow astronomers to mitigate the blurring of ground-based astronomical observations by the atmosphere of Earth \citep{Foy1985,Davies2012}. The first systems coupled optical lasers with infrared astronomical instruments, for which the adaptative optics corrections are less demanding \citep{Tyson1994,DOrgeville2016}. The integral field spectrograph {\it Multi-Unit Spectroscopic Explorer} \citep[{\muse};][]{Bacon2010} is mounted on the Unit Telescope 4 (UT4) of the European Southern Observatory's (ESO) Very Large Telescope (VLT) at the observatory of Cerro Paranal. It has been one of the very first astronomical instruments operating at visible wavelengths to be paired with an optical laser guide-star system: the four-laser guide-star facility \citep[4LGSF;][]{BonacciniCalia2014}, which is part of UT4's Adaptive Optics Facility \citep[AOF;][]{Arsenault2013}.\\

The fact that Rayleigh- and Mie-scattered laser guide-star photons can contaminate astronomical observations is well known by the astronomical community \citep[]{Delplancke1999,ChuecaUrzay2000,Summers2003,Hayano2003,Amico2010}. It is the combination of \muse\ with the 4LGSF that first revealed that astronomical observations can also be contaminated by Raman-scattered laser guide-star photons \citep[][]{Vogt2017b}. Specifically, inelastic Raman scattering physics implies that the excitation of the first vibrational mode $\nu_{1\leftarrow0}$ of the most abundant air molecules (N$_2$ and O$_2$) already redshifts laser guide-star photons by up to 2700\,cm$^{-1}\approx1000$\,\AA. In the wake of its observation at the VLT, this \textit{Raman signal} was also reported at other professional astronomical observatories \citep[][]{Lombardi2017, Marin2018, Kawaguchi2018,Lombardi2022}, and a series of specific follow-up observations were undertaken at Cerro Paranal to characterize it. These observations revealed that in the case of UT4 and the off-axis launch configuration of the 4LGSF, dust particles on the primary telescope mirror are primarily responsible for redirecting Raman-scattered laser photons inside the {\muse} field-of-view \cite{Vogt2018c}. \\

 Inserting notch filter(s) in the optical path of affected astronomical instruments -- to block the contaminating Raman-scattered photons -- would drastically reduce the spectral coverage (and thus the scientific usefulness) of these instruments. While this choice could remain meaningful for narrow-wavelength or low-resolution spectrographs, it does not make sense for most spectrographs used by the astronomical community. Specific measures \citep[e.g., keeping the telescope optics as clean as possible, see][]{Vogt2018c} can be taken to mitigate the impact of the contamination. In practice, instruments contaminated (at any level) by Raman-scattered photons -- be they emitted by their own supporting laser guide-star systems, or by a near-by one at dense observing sites -- will inevitably require their data to be cleaned-up in a dedicated post-processing step. For example, a \textit{Raman cleanup} step is now part of the official {\muse} data reduction pipeline \citep{Weilbacher2020}. The necessity of such a step for the data reduction pipeline of {\harmoni} \citep[][]{Thatte2021}, a first generation instrument of ESO's Extremely Large Telescope \citep{DeZeeuw2014}, is currently being assessed (J.\,Richard, private communication). Removing the Raman signal from contaminated data requires, in part, a detailed catalogue of molecular transitions and associated Raman shifts. Such a catalogue would benefit any science case requiring high spectral fidelity, provided that it is accurate enough.\\

The Raman signature of the most abundant air molecules is comprised of a series of dense line forests with intensity variations of several orders of magnitude. A higher-resolution spectrograph and/or better signal-to-noise observation will thus reveal a more spectrally complex Raman signal. This fact becomes evident when comparing, for example, the observations from the {\muse}, {\fors} \citep{Appenzeller1998}, and \textit{\'Echelle SPectrograph for Rocky Exoplanets and Stable Spectroscopic Observations} \citep[ESPRESSO;][]{Pepe2010,Pepe2013,Pepe2021} instruments (with respective spectral resolutions $R=\lambda/\Delta\lambda$ of $\sim$2500, $\sim$2900, and $\sim$140'000) of the $\nu_{1\leftarrow0}$ rotational-vibrational forests for the \nn\ and \oo\ molecules \citep{Vogt2017b, Vogt2019a}. It was thus suggested that Raman-scattered laser photons could also have positive uses for astronomical observatories: for example, to characterize the spectral accuracy of astronomical spectrographs via the exact same light path as science photons  \citep{Vogt2019a}, a feat unattainable with traditional calibration sources. \\

In this article, we present the deepest (i.e., highest signal-to-noise) and most accurate spectrum of the pure-rotational and rotational-vibrational spectrum of the atmosphere recorded with an astronomical spectrograph to date. This spectrum, acquired with {\espr}, is intended to provide astronomers with an exhaustive catalogue of Raman lines, associated to the most abundant air molecules, that can contaminate their observations.\\

The article is structured as follows. We provide an exhaustive description of our non-standard observations in Sec.~\ref{sec:obs}, including our data acquisition and reduction procedures. The molecular lines visible in the resulting spectrum are discussed in Sec.~\ref{sec:deep}. The discovery of distinct lasing wavelengths for the different laser beams of the 4LGSF is presented in Sec.~\ref{sec:stability}. We summarize our conclusions in Sec.~\ref{sec:conclusions}. All wavelengths are quoted in vacuum. Throughout the article, we express laser photon energies and energy differencies in different units (\AA, fm, cm$^{-1}$, m\,s$^{-1}$, Hz) following the usual practices of the applicable sub-field (astrophysics, molecular physics, or laser physics). Unless specified otherwise, the conversion between these units is made with respect to $\lambda_{\rm \lbeam}=5891.59120$\,\AA. We refer to $N$ as the end-over-end rotational quantum number of a given molecule, and $J$ its total angular momentum quantum number. We follow the usual molecular branch-naming conventions with $O\equiv(\Delta J=-2)$, $P\equiv(\Delta J=-1)$, $Q\equiv(\Delta J=0)$, $R\equiv(\Delta J=+1)$ and $S\equiv(\Delta J =+2)$.


\section{Observations}\label{sec:obs}
\subsection{Atmospheric conditions}\label{sec:atmo-state}
\begin{figure*}[htb!]
\centerline{\includegraphics[scale=0.5]{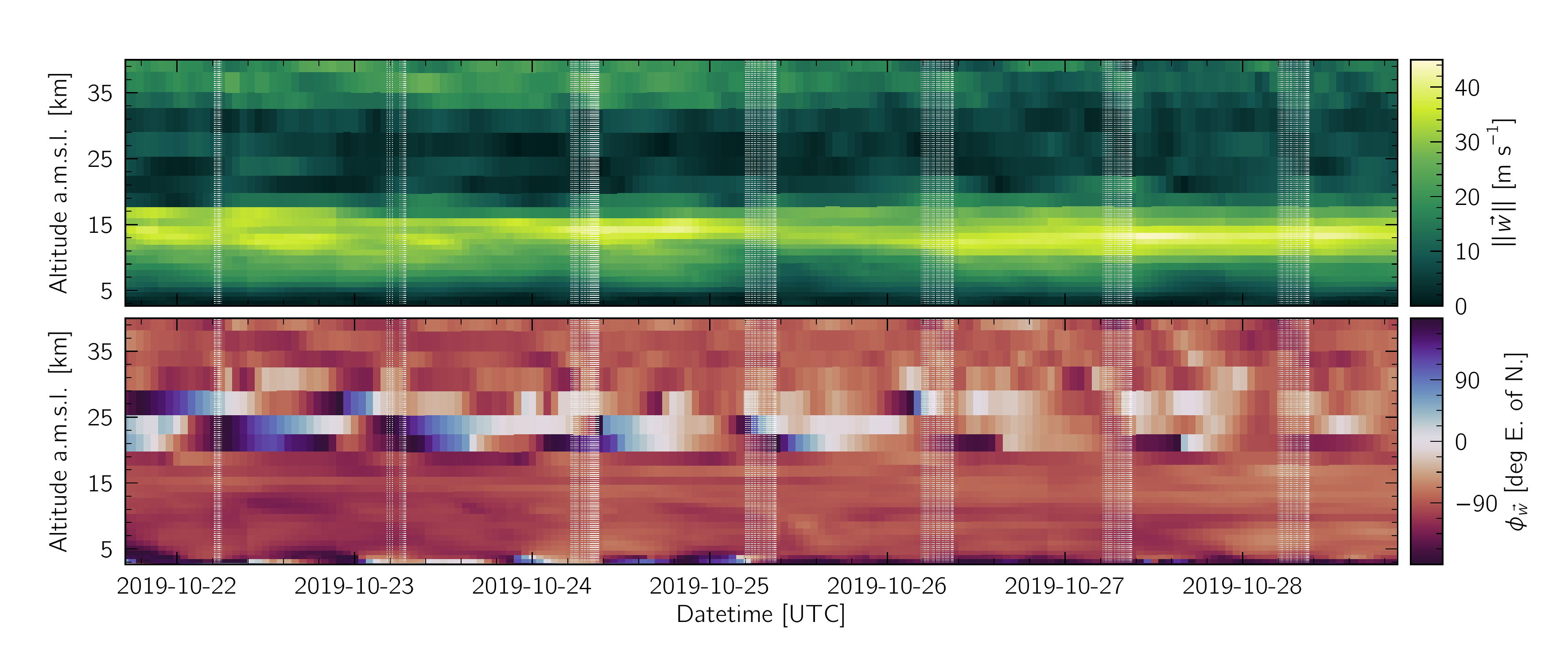}}
\caption{Top: horizontal wind speed $\left\Vert\vec{w}\right\Vert$ above Cerro Paranal over the course of the ESPRESSO observing run 4104.L-0074(A), derived from the ECMWF ERA5 re-analysis dataset. The altitude axis is cropped at the altitude of UT4 (2648\,m amsl). The white vertical lines denote the start of every 175 ESPRESSO exposures acquired over the course of the observing run. Bottom: idem, but for the wind azimuth $\phi_{\vec{w}}$ in degrees East of North ($0^{\circ}\equiv$ wind coming from the North; $90^{\circ}\equiv$ wind coming from the East). These stable conditions with dominant winds from the West are typical of this geographical location.}\label{fig:era5}
\end{figure*}

The data presented in this article were acquired in ESO's period 104 under the calibration program Id. 4104.L-0074(A) (P.I.: Vogt), for which a total of 24\,h of dark time with {\espr} on VLT UT4 were awarded. The observations took place between 21 October 2019 and 27 October 2019 (included), specifically between the half-night mark and the rise of the Moon in the East. All the observations were performed under \textsc{clear}\footnote{The official ESO definition for a \textsc{clear} sky transparency is as follows: Less than 10\% of the sky (above 30 degrees elevation) covered in clouds, transparency variations under 10\%.} sky conditions, with no visible clouds reported by the observatory's weather officer. We present in Fig.~\ref{fig:era5} the horizontal wind speed $\left\Vert\vec{w}\right\Vert$ and azimuth $\phi_{\vec{w}}$ above Cerro Paranal over the course of the entire observing run, derived from the ERA5 re-analysis dataset from the European Center for Medium-Range Weather Forecast \citep[ECMWF;][]{Hersbach2022}. Specifically, we used the Python library \textsf{cdsapi} \footnote{\textsc{cdsapi} is the official Copernicus Climate Change Service (C3S) Climate Data Store (CDS) API client. https://cds.climate.copernicus.eu/api-how-to} to request ERA5 re-analysis wind information for 37 pressure levels interpolated at the location of Cerro Paranal (24.62684$^{\circ}$ South, 70.40453$^{\circ}$ West) over the duration of the observing run. We converted the geopotential height $h$ of every data point to geometrical altitudes above mean sea level (amsl) $H$ via the following ECMWF prescription applicable to ERA5 products in the GRIB-1 data format:
\begin{equation}
H = R_e\frac{h}{R_e-h}
\end{equation}
with $R_e=6367.47$\,km.\\

The wind speed at ground level never reached more than 10\,m\,s$^{-1}$ during the observations. At high altitudes, a persistent jet stream current coming from the West was present at an altitude of 12--15\,km amsl every night of the observing run, with horizontal speeds in the range of 20--40\,m\,s$^{-1}$ that are typical for this geographical location \citep[][]{Cantalloube2020}. \\

\subsection{Data acquisition}\label{sec:acq}
The performed observations were highly non-standard: they targeted the four up-link laser beams of the 4LGSF laser guide-star system of UT4, which {\espr} is not intended/designed to exploit for Adaptive Optics purposes ({\espr} is optimized for seeing-limited observations). These observations were made possible, in large, thanks to the Laser Pointing Camera \citep[LPC;][]{BonacciniCalia2014a, Centrone2016}, which allows the telescope operator to use the laser guide-star system independently from the astronomical instrument. Since {\espr} is entirely oblivious to the 4LGSF, no information regarding the status of the laser guide-star system is recorded in the header of the {\espr} raw data files. This information was thus recorded manually for each of the 175 exposures acquired during the observing run. We provide it in Table~\ref{tbl:obs-log} in the Appendix~\ref{app:obslog}, to enable re-analysis by interested readers.\\

All our observations used the \texttt{singleHR} mode of {\espr} with the \texttt{2x1\_SLOW} detector readout mode. At the time, this provided the largest on-sky collecting fiber (1\,arcsec diameter on-sky) and lowest associated read-out noise \footnote{In 2021 a heavier binning scheme with 4 pixels in the spatial direction and 2 pixels in the dispersion direction was implemented for the \texttt{singleHR} mode of {\espr}.}. We targeted empty fields with no entries in the GAIA \citep{GaiaCollaboration2016a} Data Release 2 \citep[][]{GaiaCollaboration2018} and USNO-B2 \citep{Monet2003} catalogues within a radius of 10\,arcsec. This ensures a homogeneous background with a V-magnitude of $\sim$21 or higher \citep[][]{Patat2006} for our observations. The targeted fields are located in the declination range $-35^{\circ}\lesssim{\rm Dec.}\lesssim-40^{\circ}$ to minimize the speed of the field rotation when crossing the meridian. The majority of the observations were made at R.A.:~$04^{\text{h}}00^{\text{m}}29^{\text{s}}$; Dec.:~$-39^{\circ}52^\prime30^{\prime\prime}$. This field could be tracked continuously during our observing half-nights, which allowed to maximize the time spent on-source while systematically remaining in the telescope elevation range of $55^{\circ}-75^{\circ}$.\\ 

All observations were performed with the telescope strongly defocused. This allowed to spatially concentrate the emission from the laser beams that (unlike astronomical targets) are not located at ``infinity''. The defocus was achieved by offsetting the secondary telescope mirror by $+10$\, mm from its nominal position. No active optics corrections can be applied to the primary mirror of a defocused Unit Telescope at the VLT \citep{Wilson1987, Noethe1988}. Yet, as the telescope tracks, it remains important to account for the changing gravity vector to maintain a good image quality. We thus manually triggered (after temporarily re-focusing the telescope at infinity to detect stars in the field) several active optics correction cycles for the primary mirror every hour.\\ 

In regular operations, the {\espr} instrument provides an automated secondary field stabilization. This helps maximize light collection and ensure an homogeneous illumination of the fiber(s) during integration. This automated image stabilization is performed on the science target with a technical CCD \citep[TCCD;][]{Riva2012}, and complements the primary field stabilization performed by the telescope using a guide-star. For our observations, the field stabilization from both the telescope and {\espr} were disabled. The former because of the strong defocus of the telescope, and the latter because of the spatially-extended nature of the up-link laser beams.\\

\begin{figure*}[htb!]
\centerline{\includegraphics[scale=0.5]{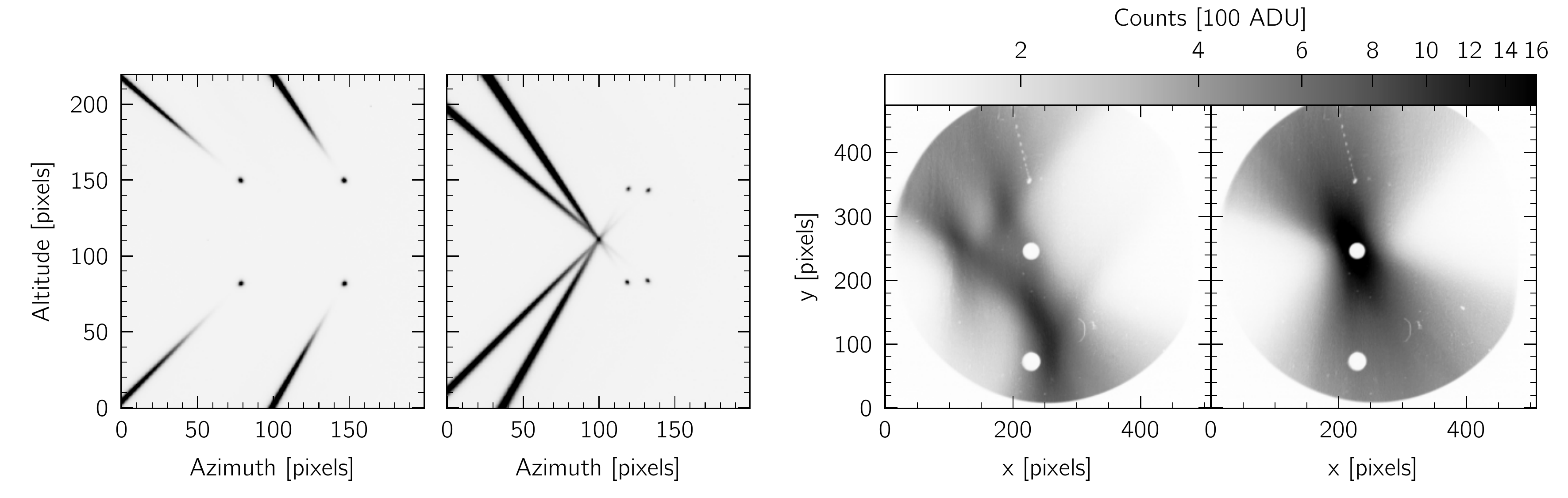}}
\caption{Left: LPC images of the 4LGSF up-link laser beams and associated guide-stars in 1) the nominal \textit{Wide-Field-Mode} asterism and 2) after manual centering of the beams onto the {\espr} fiber. The slanted perspective is a consequence of the off-axis position of the LPC, mounted on the top ring of UT4. Right: {\espr} TCCD view of the up-link laser beams before and after the finer alignment onto the {\espr} fiber A (central white dot, 1\,arcsec in diameter). Prior to the fine tuning, the individual patterns from the four beams is distinguishable, with their narrowest part corresponding to the distance in-focus ($\sim$21.5\,km along the line-of-sight, see Sec.~\ref{sec:foc-alt}). The strong spatial dependence of the observed beams' width is a direct consequence of the launch configuration for the 4LGSF lasers. This implies that the return flux seen by {\espr} is strongly sensitive to the fine alignment of the laser beams over the {\espr} fibers. The second white dot towards the bottom of the TCCD images corresponds to fiber B.}\label{fig:LPC_vs_TCCD}
\end{figure*}

The narrowest ($\equiv$ in focus) part of the 4LGSF laser beams were moved, one-at-a-time, to the on-sky location of the {\espr} fiber ``A'' as part of the acquisition sequence (see Fig.~\ref{fig:LPC_vs_TCCD}). A smooth drift of the laser beams over time required small re-adjustments of their positions with respect to the fiber every $\sim$20\,min. This drift, and the regular manual corrections performed sequentially for each laser beam, lead to:
\begin{enumerate}
\item fluctuations in the observed flux of the Raman lines of up to a factor of 2 amongst the different exposures from fiber A, and
\item a time-varying mixing ratio between the fluxes of the individual laser guide-star units (LGSUs) over both {\espr} fibers.
\end{enumerate}

For legacy and reproduceability purposes, our detailed acquisition sequence is included in Appendix~\ref{sec:acq-seq}. In every exposure, the fibre ``B'' of {\espr} was open to the sky. Its orientation is fixed with respect to fiber A in the image plane. Given the fact that {\espr} does not have a field de-rotator, fibre B did not always collect light from the same area ($\equiv$ height) of the laser beams, which are fixed in the Altitude-Azimuth frame of the telescope.\\

The observations are comprised of four main sets of exposures:
\begin{itemize}
\item $\mathfrak{set\ 1}$: $92\times600$\,s exposures acquired with the four LGSUs propagating, intended to assemble the deep spectrum of the pure-rotational and rotational-vibrational Raman signature of air.
\item $\mathfrak{set\ 2}$: $3\times1200$\,s exposures with no laser propagation, to assemble a sky reference spectrum.
\item $\mathfrak{set\ 3}$: $1\times4\times300$\,s and $6\times4\times120$\,s exposures, a series of quadruplet exposures acquired at the end of each half night to assess the spectral consistency of the different LGSUs. For each quadruplet (one per night), we cycled through all four LGSUs sequentially, enabling a single beam per exposure.
\item $\mathfrak{set\ 4}$: $11\times1$\,s exposures with only 1 laser beam (LGSU3) propagating, and an additional telescope focus offset of $+1$\,mm (for a total focus offset of $+11$\,mm). They are meant to assemble an unsaturated spectrum of the main laser line.
\end{itemize}
All the other exposures listed in Table~\ref{tbl:obs-log} that do not belong to any of these four sets were used for alignment, adjustment, and fine-tuning purposes.\\

With the exception of the night of 21 October 2019, a set of manual laser frequency comb (LFC) wavelength calibrations were triggered immediately before the start of the observations. Together with the daily calibrations triggered immediately after the end of the nights, these ensure than no scientific exposure was acquired more than $\sim$2.5\,hr apart from an LFC reference calibration exposure. LFCs are state-of-the-art, extremely precise and accurate wavelength calibration sources \citep{Frank2018,Zhao2020,Milakovic2021}. These devices can generate dense forests of narrow lines with even intensities and a uniform line spacing, that can be matched to the resolution and wavelength range of high-resolution spectrographs like {\espr}. The frequencies of the LFC lines can be locked to an atomic clock, ensuring a precision and accuracy of up to one part in $10^{11}$. At the time of our observations, {\espr}'s LFC was formally under commissioning, but nonetheless operational.

\subsection{Sampled altitude range}\label{sec:foc-alt}
We do not have a direct measure of the altitude at which the recorded Raman signal originated. However, we can obtain an estimate of the distance (along the line-of-sight) that is in-focus during the observations from basic telescope optics. UT4 has an effective focal length $f_\mathrm{eff}=121.561$\,m. Its secondary mirror has a radius of curvature $r=4'553.1$\,mm \citep[][]{Harel2016}. According to the VLT optical layout, the Nasmyth focus lies $a=9'896+6'800=16'696$\,mm away from the surface of the secondary mirror. Treating the secondary mirror as a spherical rather than hyperbolic surface (for the sake of simplicity) implies that the virtual focal plane of the primary mirror lies $a^{\prime}\approx ra/(r+2a)=2'003$\,mm behind the surface of its secondary mirror \cite{Rupert1918}. When the telescope is focused at infinity, the image is formed at a distance $l=f_\mathrm{eff}$. Offsetting the secondary mirror by $\Delta x=+10$\,mm changes the effective focal length of the telescope to:
\begin{equation}
    f_\mathrm{eff}^{\star} = f_\mathrm{eff}\cdot\frac{a^{\prime}}{a}\cdot\frac{a+\Delta x}{a^{\prime}-\Delta x}=122.24\,\text{m.}
\end{equation}
The camera being fixed, its distance $l$ along the optical path translates to an object plane located at a distance $D$ along the line-of-sight given by:
\begin{equation}
    D=\left(\frac{1}{f_\mathrm{eff}^{\star}}-\frac{1}{l}\right)^{-1}=21.8\,\text{km.}
\end{equation}
This analytical estimate was verified using a complete model of UT4, its Coud\'e train \citep[][]{Cabral2019}, and the {\espr} instrument, implemented in a state-of-the-art ray-tracing optical design software. Offsetting the secondary mirror by $+10$\,mm in the model placed the object plane at a distance of $D=21.5$\,km along the line-of-sight (A. Cabral, private communication). For a given UT4 pointing altitude $\psi$, this corresponds to a geometric altitude above mean sea level $H_{\rm obs}=D\sin\psi+H_{\rm UT4}$ with $H_{\rm UT4}=2.648$\,km. The median value of $H_{\rm obs}$ for the $\mathfrak{set\ 1}$ exposures is 23.0\,km amsl.\\

\begin{figure}[htb!]
\centerline{\includegraphics[scale=0.4]{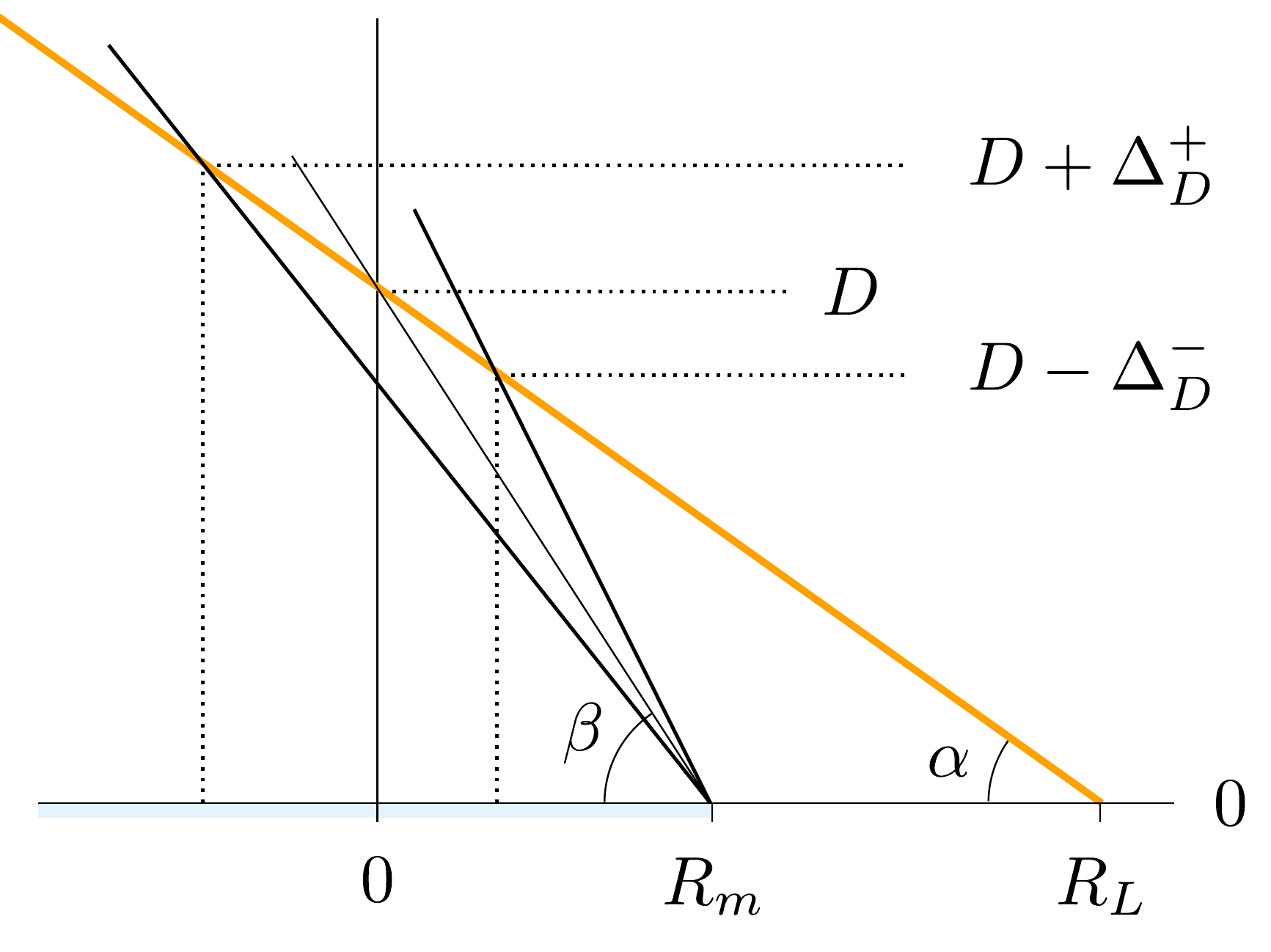}}
\caption{Simple schematic of the UT4 primary mirror (with a diameter $R_m$) and one 4LGSF laser beam (not to scale). The telescope is assumed to be focused at a distance $D$ along the line-of-sight. The finite aperture size of the {\espr} fibers and the slow, regularly-and-manually-corrected drift of the laser beams imply that emission from a range of distances along the line-of-sight will be collected. The maximum range limits $\Delta_D^-$ and $\Delta_D^+$ are set by the optical paths passing closest (when hitting the UT4 mirror) from the laser launch telescopes.}\label{fig:alt-range}
\end{figure}

The finite aperture size of the {\espr} fibers (1\,arcsec in diameter) coupled to the slow, regularly-and-manually-corrected drift of the laser beams imply that emission from a range of distances $[D-\Delta_D^-;D+\Delta_D^+]$ along the line-of-sight will be sampled in the observations. We estimate this range using the simple model depicted in Fig.~\ref{fig:alt-range}, with $R_m=4.1$\,m the radius of the UT4 primary mirror, and $R_L=5.51$\,m the distance between the optical axis of UT4 and the location where the 4LGSF laser beams would cross the mirror plane \citep[if extended backward from the launch telescope, see the Appendix in][]{Vogt2017b}. The maximum range of distances sampled by the {\espr} fiber A, when focused at a distance $D$ along the line-of-sight, is set by the optical paths passing the closest to any of the four LGSUs' launch telescopes. From basic trigonometry, given a fiber diameter of 1\,arcsec on-sky and an estimated motion of the laser beams up to $\sim1$\,arcsec beyond this diameter, we have:
\begin{equation}
    D\pm\Delta_D^\pm =\frac{(R_L-R_m)\tan\alpha\tan\beta^\mp}{\tan\beta^\mp-\tan\alpha} 
\end{equation}
with $\tan\alpha=D/R_L$, $\beta^\mp=\beta\mp1.5$\,arcsec, and $\tan\beta=D/R_m$. With a telescope focus set at $D=21.5$\,km, we thus estimate that $\Delta_D^-\approx 2.2$\,km and $\Delta_D^+\approx2.7$\,km, such that the range of observed distances along the line-of-sight in the data is [19.3\,km; 24.2\,km]. The corresponding geometric altitude range for every {\espr} exposure acquired during the observing run is presented in Fig.~\ref{fig:wlos}: it lies systematically within the interval [18\,km; 26\,km] amsl.

\subsection{4LGSF state}

All four LGSUs of the 4LGSF were used in their nominal mode for the majority of the exposures acquired during the observing run. In this mode, the emitted light is comprised of 18\,W in a central component designed to excite the D$_2$a sodium transition, and $2\times2$\,W in symmetric side-lines located $\pm1.713$\,GHz on either side \citep{Vogt2019a}. The lower-energy side-line is designed to excite the D$_2$b sodium line, and to recover atoms lost to the $3^2S_{1/2}$ $F=1$ state \citep[][]{Holzlohner2010}.\\

The 4LGSF 589\,nm$\approx$509\,THz laser sources (one per LGSU) are based on a frequency-doubled fibre amplifier. The output emission wavelength of the 1178\,nm$\approx$255\,THz master oscillators (seed lasers) of the 589-nm laser sources are actively stabilized to twice the sodium resonance wavelength via wavelength meters (one per LGSU) with a specified accuracy of better than $\pm10$\,MHz \citep{BonacciniCalia2010}. The wavelength meters exhibit a passive stability of about 80\,kHz\,h$^{-1}$ (linear drift). The absolute accuracy of the wavelength meters is guaranteed by periodical calibrations against iodine-referenced helium-neon lasers (one per LGSU) which themselves have a specified absolute accuracy of better than $\pm$5\,MHz. The feedback loop which stabilizes the seed lasers to the wavelength meters is another factor contributing to the actual LGSUs emission wavelengths. This feedback loop has an accuracy of about $\pm5$\,MHz at 1178\,nm, which was determined during the development of the laser sources.\\

The four wavelength meters were calibrated every afternoon before our observing nights. However, it must be noted that the 4LGSF system was rebooted during the first half of night 6 (i.e. after the afternoon wavelength meter calibrations were executed), and no new wavelength meter calibrations could be executed before the start of the {\espr} observations.\\

For a handful of exposures, one or more of the four LGSUs were detuned for test purposes. The detuning of each LGSU is configured in the laser control software. When detuned, LGSU1, LGSU2 and LGSU4 wavelengths are (blue-)shifted by $+$5.960\,GHz at 589\,nm; LGSU3 is (red-)shifted by $-$11.057\,GHz at 589\,nm. These exposures will not be discussed in this article, but they are clearly identified in Table~\ref{tbl:obs-log}.

\subsection{Saturation-induced artefacts}\label{sec:artefacts}

\begin{figure*}[htb!]
\centerline{\includegraphics[scale=0.5]{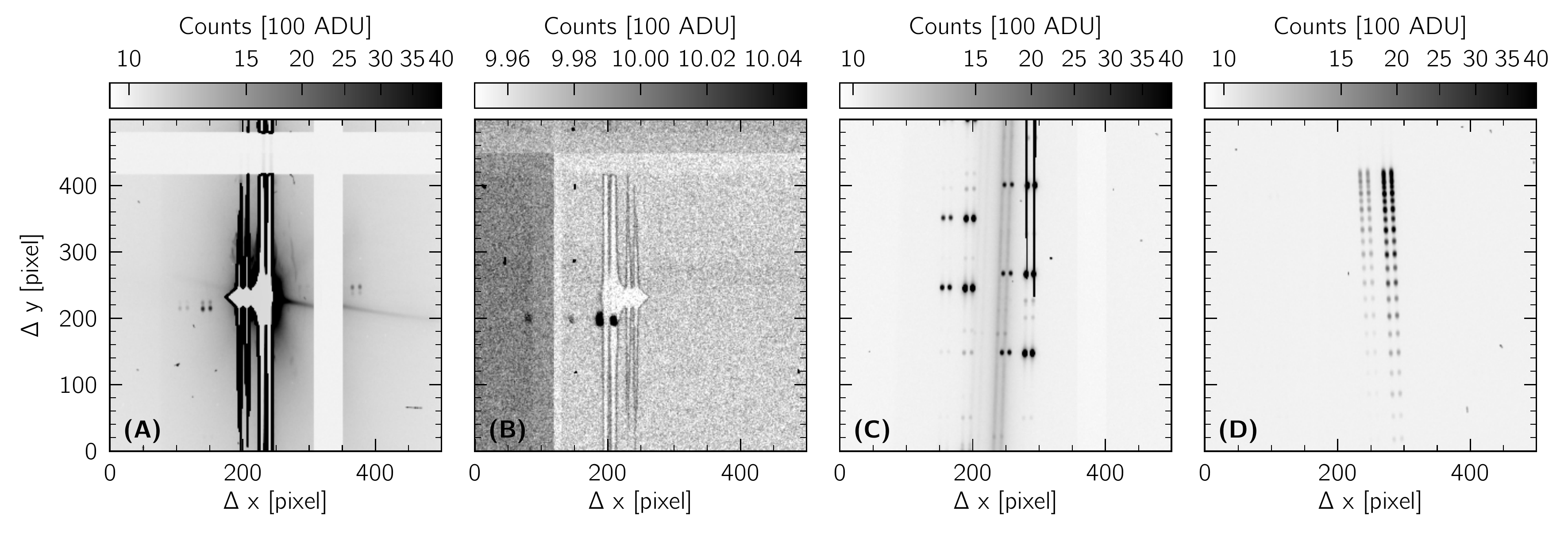}}
\caption{Subsections of a single $\mathfrak{set\ 1}$, 600\,s {\espr} raw exposure from the red CCD. (A) region of the main laser line, saturated. Strong blooming (with electrons leaking vertically across the detector) is evident, rendering this specific order un-exploitable. (B) Other regions of the detector display ghost images of the saturated region, but significantly closer to the noise level. (C) In the vicinity of the main laser line, pure rotational Raman lines (black dots) are also being contaminated by an artificial continuum emission (middle-grey emission) in addition to the blooming from the main laser line (vertical black stripes). (D) Most areas of the red detector are free from strong artefacts, but remain subject to an elevated, smooth background signal: for example, the region of the $\nu_{1\leftarrow0}$ rotational-vibrational Q-branch of the \nn\ molecule shown here.}\label{fig:ccds}
\end{figure*}

The strength of the Raman lines is several orders of magnitude ($\geq3$) fainter than the strength of the main laser line resulting from elastic Mie and Rayleigh back-scattering. For an optical CCD like that of {\espr} with a range of $\sim$65'000 ADU mounted on a telescope with a collecting area 8\,m in diameter, the main laser line already saturates in exposures as short as 1\,s (with the beam in focus). For the significantly-fainter Raman lines, on the other hand, the longer the {\espr} exposures are, the stronger the signal-to-noise ratio of the lines, and the more efficient the observations.\\

The use of 600\,s exposures for our observations is the result of the trade-off between (a) the need to enhance the signal-to-noise ratio of the Raman lines in individual exposures, and (b) limit the total number of individual exposures to minimize the read-out overheads, while (c) maintaining the degree of saturation bleeds and ghosts on the {\espr} CCD within reason. In 600\,s, the saturation of the main laser line on the red {\espr} CCD is severe nonetheless, as illustrated in Fig.~\ref{fig:ccds}. Blooming from the saturated pixels strongly contaminates pixels immediately above and below the main laser line, such that this entire spectral order is effectively lost. Other artefacts caused by the strength of the main laser line include:
\begin{itemize}
    \item bright trails due to charge transfer inefficiency,
    \item an artificial continuum in the vicinity of the main laser line,
    \item ghost images of the saturated regions caused by internal reflections, and
    \item a higher background for the entire red CCD, illuminated by scattered light.
\end{itemize}
The exquisite stability of the {\espr} instrument implies that all these artefacts do not vary between exposures, except for their respective brightness which is tied to the incoming laser flux.\\

\subsection{Data reduction}
The raw data were downloaded from the ESO Science Archive Facility and reduced using the {\espr} pipeline v2.0.0 \citep{Pepe2021} via the \textsc{EsoReflex} v2.11.0 environment \citep[][]{Freudling2013}. The data processing consisted of 1) bias and dark subtraction, 2) optimal extraction of spectral orders and creation of extracted spectra (S2D), and 3) wavelength calibration of S2D spectra using the LFC. Subtraction of the inter-order background was turned off, because it led to sub-optimal results due to the strong bleeding of the saturated laser line. We use the reduced S1D spectra from fibre A in the subsequent analysis, for which we systematically remove the barycentric Earth radial velocity (stored under the keyword \texttt{HIERARCH ESO QC BERV} in the {\espr} FITS files) correction applied by default by the {\espr} pipeline via:
\begin{equation}
\lambda = \lambda_{\rm raw}\left[\left(1+1.55\cdot10^{-8}\right)\cdot\left(1+\frac{{\tt BERV}}{299792.458}\right)\right]^{-1}
\end{equation}

The 15.3\,hr-deep {\espr} spectrum of the pure-rotational and rotational-vibrational Raman signature of our atmosphere at an altitude of $\sim$23\,km amsl is assembled using the $\mathfrak{set\ 1}$ ($92\times600$\,s), resampled, blaze-corrected, one-dimensional spectra from fibre A, in units of photo-electrons. These exposures are stacked together on a common wavelength grid using a flux weighted-average combination scheme. This choice is motivated by the fluctuations of the laser line fluxes between the different exposures (up to a factor $\sim$2, see Sec.~\ref{sec:acq}).\\

The spectral regions 5880.8\,\AA$~<\lambda<$~5899.5\,\AA\ and 5875.1\,\AA$~<\lambda<$~5875.8\,\AA, are strongly affected by the blooming of the main laser line. They are replaced with the (scaled) weighted-average spectrum assembled from the $\mathfrak{set\ 4}$  ($11\times1$\,s) exposures that show no saturation. The resulting, combined spectrum ranges from 3775\,\AA ~$\approx +9517$\,cm$^{-1}$ to 7895\,\AA ~$\approx -4307$\,cm$^{-1}$. In the assembly process, no correction of the estimated spectral shift induced by wind (see Sec.~\ref{sec:stability}) is being applied to the individual exposures. These spectral shifts have amplitudes $<20$\,m\,s$^{-1}\approx40$\,fm. This corresponds to $\lesssim5$\,\% of the Raman line widths, which are dominated by thermal broadening. Wind-induced spectral shifts thus does not limit our ability to identify individual molecular transitions in the resulting deep {\espr} spectrum.\\


\section{Results}\label{sec:results}

\subsection{The Raman signature of the atmosphere at $H_{\rm obs}=23$\,km}\label{sec:deep}

\begin{figure*}[htb!]
\centerline{\includegraphics[scale=0.5]{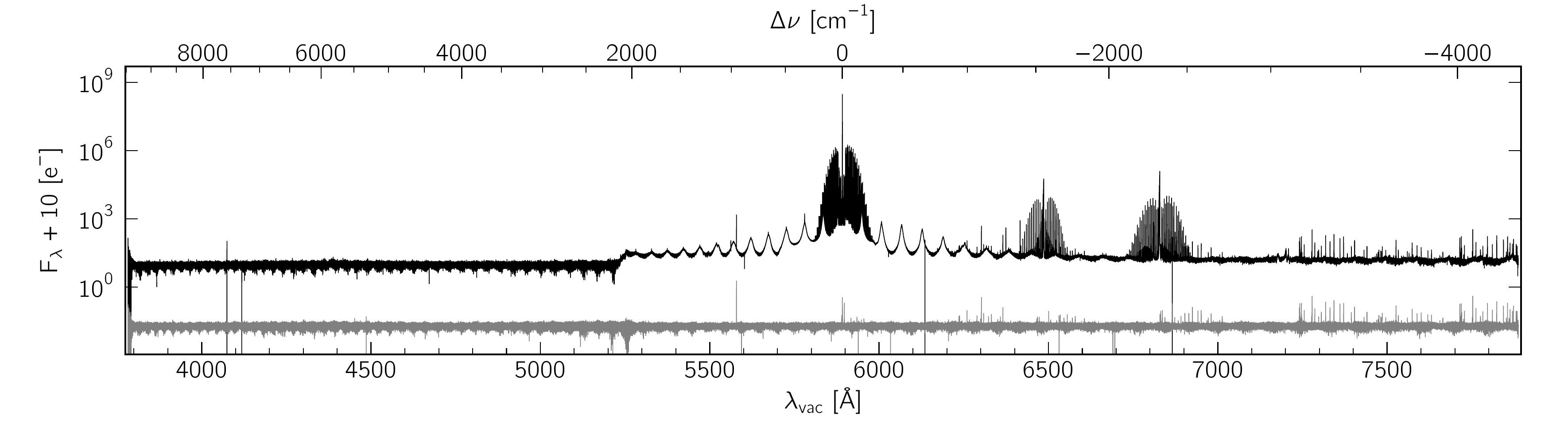}}
\caption{Weighted-average of $92\times600$\,s {\espr} exposures ($\mathfrak{set\ 1}$) of the four 4LGSF laser beams (black), with the strongly saturated regions 5880.8\,\AA$<\lambda<$5899.5\,\AA\ and 5875.1\,\AA$<\lambda<$5875.8\,\AA\ replaced by a (scaled) weighted-average combination of $11\times1$\,s exposures ($\mathfrak{set\ 4}$). The reference sky spectrum assembled from $3\times1200$\,s exposures ($\mathfrak{set\ 2}$), artificially scaled and shifted for visualization purposes, is shown in grey. The forest of pure rotational Raman lines immediately surrounding the main laser line is clearly visible, together with the $\nu_{1\leftarrow0}$ rotational-vibrational Raman forests associated to the O$_2$ (at $\sim$6485\,\AA$\approx$-1550\,cm$^{-1}$) and N$_2$ (at $\sim$6825\,\AA$\approx$-2320\,cm$^{-1}$) molecules. The jump between the blue and red CCD spectra and the extended oscillatory pattern surrounding the main laser line, clearly visible in this logarithmic view, are artefacts caused by the saturation of the main laser line.}\label{fig:full}
\end{figure*}

The 15.3\,hr-deep {\espr} spectrum revealing the pure-rotational and rotational-vibrational Raman spectrum of the atmosphere at $H_{\rm obs}=23$\,km is shown in Fig~\ref{fig:full}. The main laser line at $\Delta\nu=0$ is clearly visible, together with the three forests of:
\begin{enumerate}
    \item pure-rotational Raman lines in its immediate vicinity,
    \item $\nu_{1\leftarrow0}$ rotational-vibrational Raman lines associated with the O$_2$ molecule at $\sim6490$\,\AA, and
    \item $\nu_{1\leftarrow0}$ rotational-vibrational Raman lines associated with the N$_2$ molecule at $\sim6830$\,\AA.
\end{enumerate}

A careful comparison with the reference sky spectrum reveals $\nu_{2\leftarrow0}$ rotational-vibrational $Q$-branch lines from \oo\ at $\lambda_{\rm vac}\cong7202$\,\AA, and vibrational lines from \coo\ at $\sim$6366.1\,\AA, $\sim$6374.3\,\AA, $\sim$6416.4\,\AA, and $\sim$6425.1\,\AA\ \citep[][]{Tejeda1995}. The $\nu_{2\leftarrow 0}$ rotational-vibrational forest from \nn\ at $\Delta\nu\approx-4631.2\approx8100$\,\AA\ \citep[that can be detected with {\muse};][]{Vogt2017b}) is located beyond the spectral range of {\espr}. The complete list of 865 Raman lines (pure-rotational: 511; rotational-vibrational: 354) identified in the spectrum is presented in Table~\ref{tbl:lines} in the Appendix~\ref{app:linelist}, and accessible electronically as a VizieR \footnote{https://doi.org/10.26093/cds/vizier} catalogue \citep{Ochsenbein2000}. Several of the artefacts caused by the strong saturation of the main laser line are also readily visible in the logarithmic-scale diagram presented in Fig.~\ref{fig:full}. These include the artificial continuum with an oscillatory-like pattern, and the non-zero background illumination red-wards of 5250\,\AA~$\approx+2075$\,cm$^{-1}$.\\

A handful of emission line-like features remain unidentified at this stage. These include, in particular, 22 features located within $\pm56$\,\AA\ of the main laser line, a subset of which at least appear to be distributed symmetrically (in the Raman shift space) around the main laser line. This could suggest a molecular origin. A careful inspection of the raw {\espr} exposures also suggests that these lines are real. However, the complex and diverse nature of the artefacts caused by the strong saturation of the main laser line forbids us from stating with certainty that these 22 unidentified features are not artefacts themselves. Doing so will require a precise match of these features (listed in Table~\ref{tbl:unknown}) with specific molecular transitions, which elude us so far.

\begin{table}[htb!]
\caption{Un-indentified emission line-like features located within $\pm160$\,cm$^{-1}$ of the main laser line at $\Delta\nu=0$ in the 15.3\,hr-deep {\espr} spectrum of the 4LGSF laser beams. The vacuum wavelengths accuracy is $\sim0.01$\,\AA.}\label{tbl:unknown}
\begin{tabular}{c c p{0.4cm} c c}
\hline\hline
$\lambda_{\rm vac}$ & $\Delta\nu$ & & $\lambda_{\rm vac}$ & $\Delta\nu$\\
{[\AA]} & [cm$^{-1}$] & & {[\AA]} & [cm$^{-1}$] \\
\hline
5852.55 & +113.23 & & 5901.77 & -29.27 \\
5853.89 & +109.31 & & 5902.79 & -32.20 \\
5856.61 & +101.38 & & 5905.55 & -40.12 \\
5864.78 & +77.59 & & 5910.28 & -53.67 \\
5869.73 & +63.22 & & 5913.61 & -63.20 \\
5877.69 & +40.14 & & 5918.64 & -77.57 \\
5880.44 & +32.19 & & 5924.22 & -93.48 \\
5881.47 & +29.21 & & 5927.01 & -101.43 \\
5886.65 & +14.25 & & 5929.78 & -109.31 \\
5896.54 & -14.25 & & 5931.16 & -113.23 \\
5899.74 & -23.44 & & 5933.94 & -121.13 \\
\hline\hline
\end{tabular}
\end{table}

\subsubsection{Pure-rotational Raman lines}

\begin{figure*}[p!]
\centerline{\includegraphics[scale=0.5]{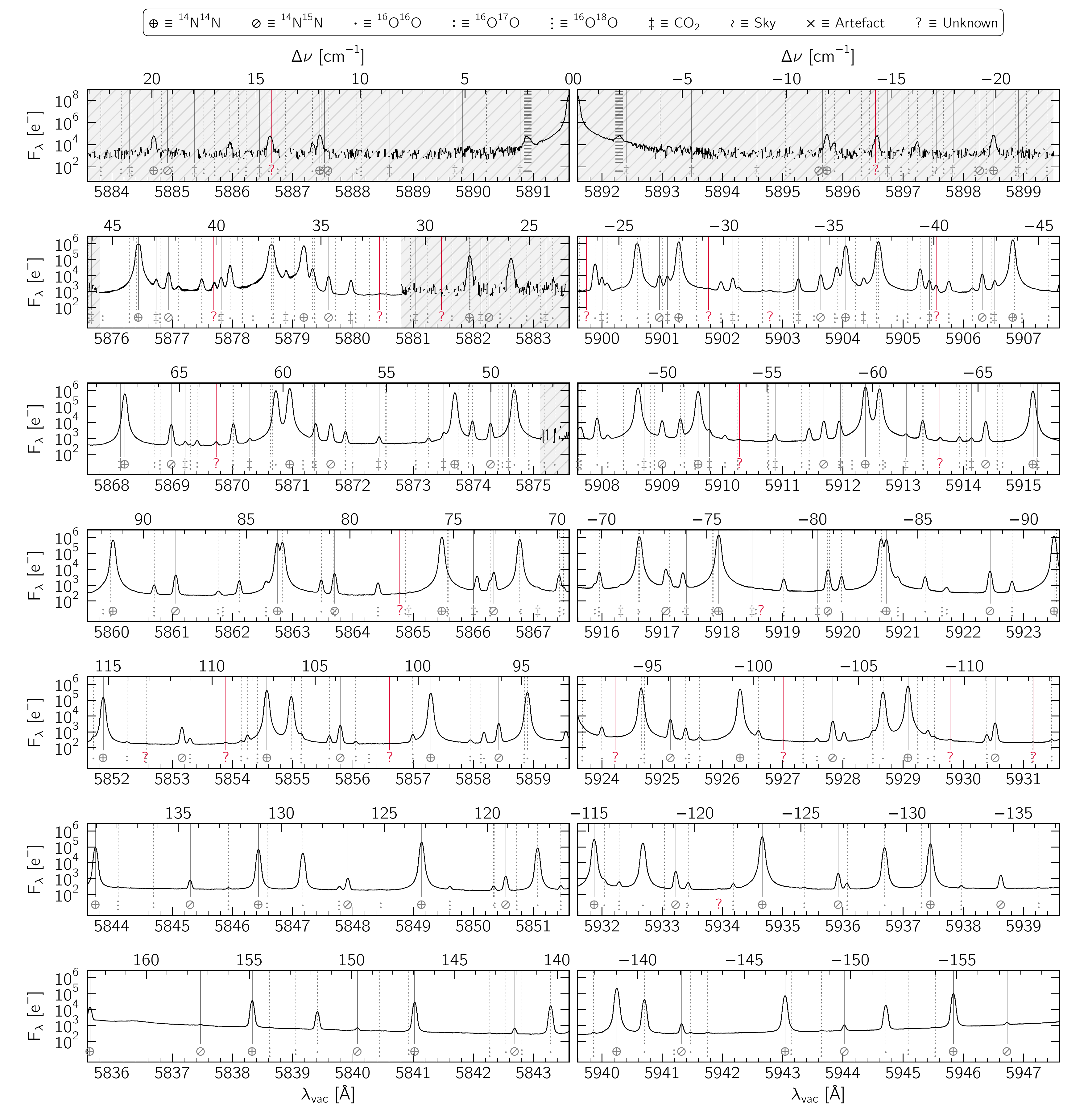}}
\caption{Pure-rotational Raman spectrum of the atmosphere at an altitude $H_{\rm obs}=23$\,km, within $\pm56$\,\AA $\approx\mp160$\,cm$^{-1}$ of the main laser wavelength. The spectrum is a weighted-average combination of the $\mathfrak{set\ 1}$ ($92\times600$\,s) {\espr} exposures. The spectral regions 5880.8\,\AA~$<\lambda<$~5899.5\,\AA\ and 5875.1\,\AA~$<\lambda<$~5875.8\,\AA\ (highlighted with a dashed-grey background) are assembled from a weighted-average combination of the $\mathfrak{set\ 4}$ ($11\times1$\,s) exposures. The theoretical location of pure-rotational transitions from  main components of dry air are marked and tagged accordingly (\nn: $\oplus$ | \nnp: $\oslash$ | \oo: $\cdot$ | \oop: $\vdotdot$ | \oopp: $\vdots$ | \coo: $\ddagger$). Sky lines are tagged with the symbol $\wr$. Artefacts caused by the strong saturation of the main laser line are tagged with the symbol $\times$. A series of unidentified line-like features, for which a Raman origin is plausible but could not be established, are tagged with red question marks.}\label{fig:rot}
\end{figure*}

Fig.~\ref{fig:rot} presents a detailed view of the spectral region within $\pm56$\,\AA\ ($\approx\mp160$\,cm$^{-1}$) of the main laser line, which reveals the densest part of the pure-rotational Raman line forest. Molecular lines associated to the most abundant molecules of air, including their isotopes, are clearly detected. These include: \nn, \nnp, \oo, \oop, \oopp, and \coo, for which we identify 511 pure-rotational lines altogether, over the spectral range [+250\,cm$^{-1}$; -265\,cm$^{-1}$]$\approx$[5806\,\AA; 5985\,\AA]. By comparison, the notch filters of {\muse} (one per adaptive optics observing mode) are blocking a spectral range of:
\begin{itemize}
\item $\approx$[5800\,\AA; 5970\,\AA] in WFM-AO-N mode,
\item $\approx$[5750\,\AA; 6010\,\AA] in WFM-AO-E mode, and 
\item $\approx$[5780\,\AA; 6050\,\AA] in NFM-AO-N mode \citep[][]{Weilbacher2020}.
\end{itemize}

It was identified during the commissioning of {\espr} that the spectral resolution of this instrument is sufficient to resolve the fine-structure Raman lines from {\oo} \citep[][]{Vogt2019a}, which has a $^3\Sigma$ electronic ground state \citep[][]{Yu2012}. This implies that unlike {\nn}, treating the {\oo} molecule as a non-rigid diatomic rotator is not sufficient to identify all its Raman lines present in the 15.3\,hr-deep {\espr} spectrum. Instead, we rely on an extensive Hamiltonian model constrained by microwave, THz, infrared, visible and ultraviolet transitions of all six oxygen isotopologues to do so \citep[][]{Drouin2012, Drouin2013, Yu2014}. The specific pure-rotational line sets visible for the \oo, \oop, and \oopp\ molecules, identified using this model, are listed in Table~\ref{tbl:rot-o2} (see Appendix~\ref{app:linelist} for details).\\

\begin{table}[htb!]
\caption{Pure-rotational line sets of the \oo\ molecule and its isotopes identified in the 15.3\,hr-deep {\espr} spectrum of the 4LGSF laser beams. The maximum value of $N$ up to which the lines of a given set could be identified is specified explicitly.}\label{tbl:rot-o2}
\begin{tabular}{l c c c}
\hline\hline
Molecule & $^{\Delta N}\Delta J$ & $(J, N)$ & $\max(N)$ \T\B\\
\hline
\oo & $^SS$ & $(N, N)$ & 41\T \\ 
\oo & $^SR$ & $(N, N)$ & 19 \\ 
\oo & $^SR$ & $(N+1, N)$ & 23 \\
\oo & $^QP$ & $(N+1, N)$ & 21 \\
\oo & $^QR$ & $(N-1, N)$ & 21 \\
\oop & $^SS$ & $(N, N)$ & 26 \\ 
\oop & $^SR$ & $(N, N)$ & 9 \\ 
\oop & $^SR$ & $(N+1, N)$ & 12 \\ 
\oopp & $^SS$ & $(N, N)$ & 27 \\ 
\oopp & $^SR$ & $(N, N)$ & 11 \\ 
\oopp & $^SR$ & $(N+1, N)$ & 20\B \\ 
\hline\hline
\end{tabular}
\end{table}

\subsubsection{Vibrational-rotational Raman lines}
\begin{figure*}[htb!]
\centerline{\includegraphics[scale=0.5]{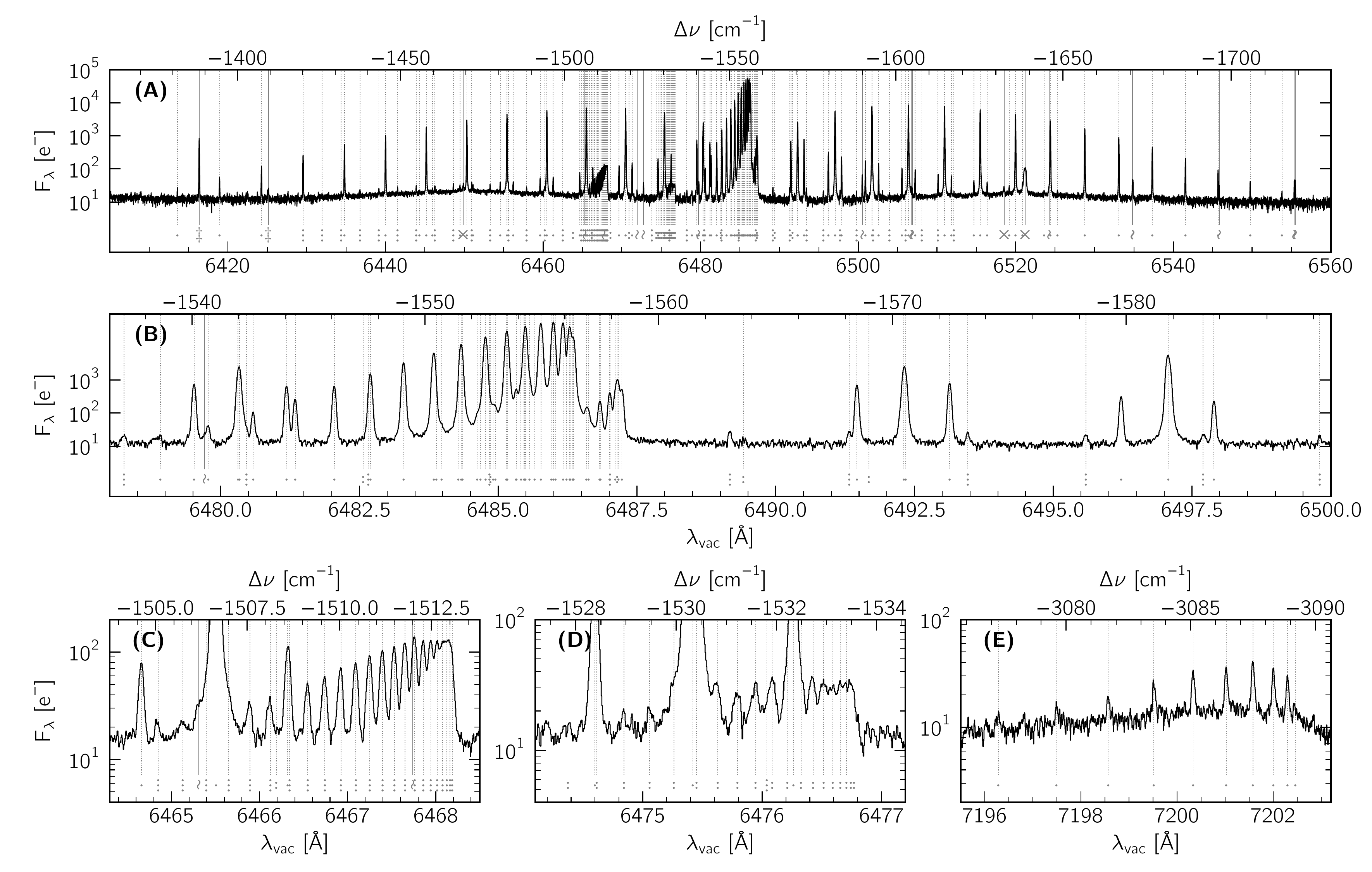}}
\caption{(A) $\nu_{1\leftarrow0}$ vibrational-rotational forest for the \oo, \oop, and \oopp\ molecules. Two vibrational transitions from \coo\ are also visible at $\lambda=6416.4$\,\AA\ and 6425.1\,\AA. (B) Detailed view of the Q-branch for the \oo\ molecule. The scaled-and-shifted feature at $\sim$6487.2\,\AA\ is a fine-structure component related to the $^3\Sigma$ electronic ground state of {\oo}, as are the symmetric side lines surrounding the $^SS\ (1,1)$ and $^SS\ (3,3)$ transitions. (C) Detailed view of the Q-branch of the \oopp\ molecule. (D) Detailed view of the Q-branch of the \oop\ molecule. (E) Detailed view of the $\nu_{2\leftarrow0}$ Q-branch of the \oo\ molecule. The theoretical position of individual Raman lines, sky lines, and artefacts are marked using vertical grey lines. Each is tagged using the symbols introduced in Fig.~\ref{fig:rot}.}\label{fig:rotvib-o2}
\end{figure*}

\begin{figure*}[htb!]
\centerline{\includegraphics[scale=0.5]{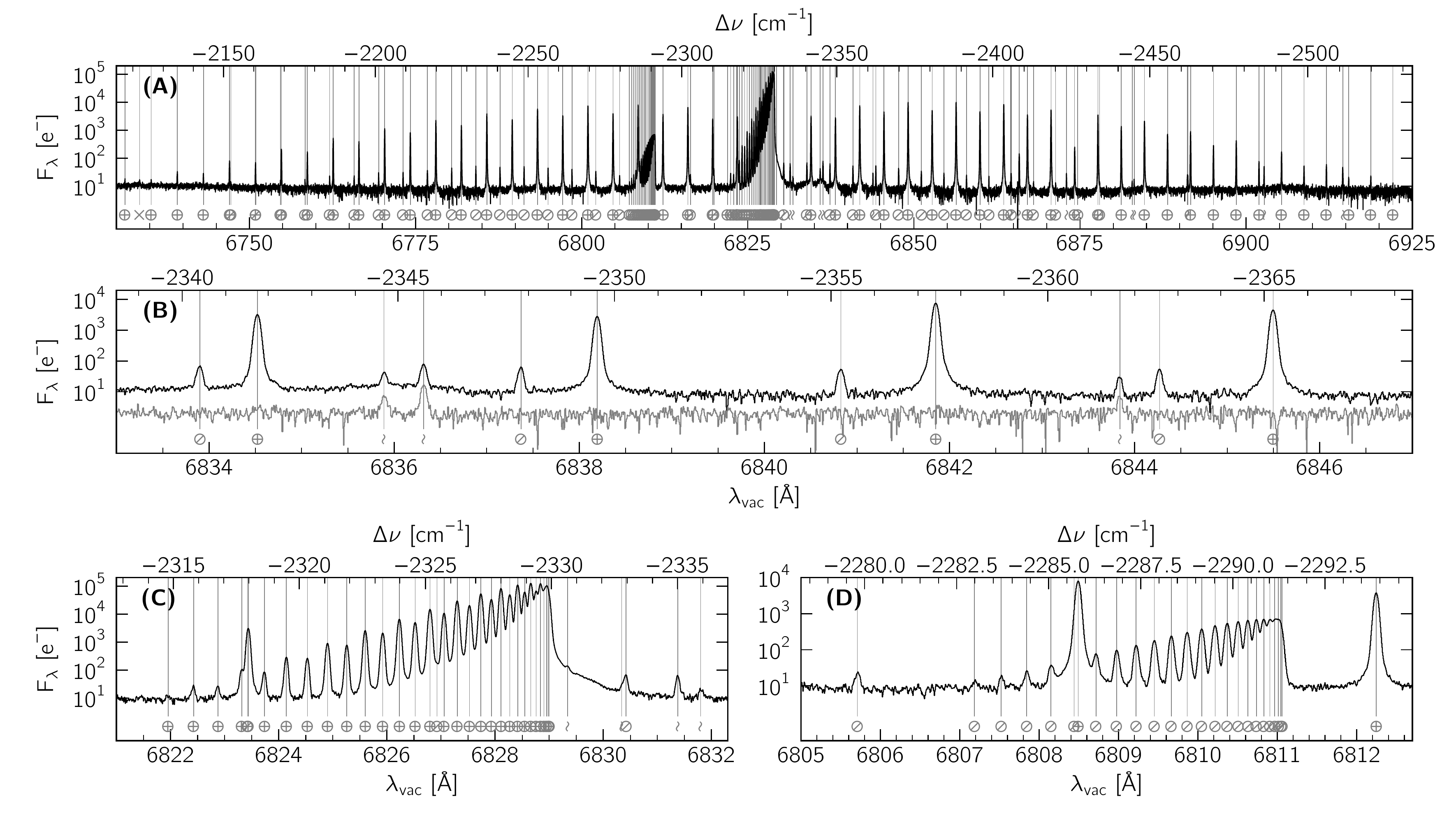}}
\caption{(A) $\nu_{1\leftarrow0}$ vibrational-rotational forest for the {\nn} and {\nnp} molecules. (B) Detailed view of a subset of Raman and sky lines. Sky lines can be unambiguously identified via their presence in the reference sky spectrum (shown in grey, scaled and shifted for visualization purposes). (C) Detailed view of the Q-branch of the \nn\ molecule. (D) Detailed view of the Q-branch of the \nnp\ molecule. In every panel, the theoretical position of individual Raman lines, sky lines, and artefacts are marked using vertical grey lines. Each is tagged using the symbols introduced in Fig.~\ref{fig:rot}.}\label{fig:rotvib-n2}
\end{figure*}

Figs.~\ref{fig:rotvib-o2} and \ref{fig:rotvib-n2} present the $\nu_{1\leftarrow0}$ rotational-vibrational forests of the O$_2$ and N$_2$ molecules, respectively. During the commisisoning of {\espr}, only the $Q$ branch of {\oo} and {\nn} was detected in a single 900\,s exposure \citep[][]{Vogt2019a}. The 15.3\,hr-deep spectrum reveals, in addition, the $Q$ branch from {\oop}, {\oopp}, and {\nnp}, as well as the $O$ and $S$ branches from {\oo}, {\oopp}, {\nn}, and {\nnp}.\\

Rotational-vibrational fine-structure lines related to the $^3\Sigma$ electronic ground state of \oo\ are detected in all the $Q$, $O$, and $S$ branches of this molecule. We list in Table~\ref{tbl:rotvib-o2} the specific sets of rotational-vibrational molecular transitions from the {\oo} molecule and its isotope that can be identified in the spectrum. For \oo, we note that the first lines (corresponding to $N=1$) of the sets $\Delta N=\text{Q}$, $\Delta J=R | P$ are not visible in the {\espr} spectrum. This is consistent with these specific transitions having zero intensity \citep[][]{Hill1979, Hill1983}. We also note that lines from the sets $^SS\ (N, N-1)$ and $^OO\ (N, N-1)$ become increasingly blended with lines from the sets $^SS\ (N,N)$ and $^OO\ (N,N)$ with increasing $N$ values. The presence of these sets is thus only inferred from their $N=1$ and $N=3$ line, respectively (visible in panel B of Fig.~\ref{fig:rotvib-o2}).

\begin{table}[tb!]
\caption{Same as Table~\ref{tbl:rot-o2}, but for the rotational-vibrational line sets of \oo\ and its isotopes.}\label{tbl:rotvib-o2}
\begin{tabular}{l c c c c}
\hline\hline
Molecule & vibrational level& $^{\Delta N}\Delta J$ & $(J, N)$ & max($N$)\B\T \\
\hline
\oo & ${1\leftarrow0}$ & $^QQ$ & $(N, N)$ & 33\T \\
\oo & ${1\leftarrow0}$ & $^QP$ &  $(N, N)$ & 16 \\
\oo & ${1\leftarrow0}$ & $^QP$ &  $(N+1, N)$ & 16 \\
\oo & ${1\leftarrow0}$ & $^QR$ & $(N, N)$ & 16 \\
\oo & ${1\leftarrow0}$ & $^QR$ & $(N-1, N)$ & 16 \\
\oo & ${1\leftarrow0}$ & $^QS$ & $(N-1, N)$ & 3 \\
\oo & ${1\leftarrow0}$ & $^QO$ & $(N+1, N)$ & 3 \\
\oo & ${1\leftarrow0}$ & $^SS$ & $(N, N-1)$ & 1 \\
\oo & ${1\leftarrow0}$ & $^SS$ & $(N, N)$ & 29 \\
\oo & ${1\leftarrow0}$ & $^SR$ & $(N, N)$ & 15 \\
\oo & ${1\leftarrow0}$ & $^SR$ & $(N+1, N)$ & 15 \\
\oo & ${1\leftarrow0}$ & $^OO$ & $(N, N-1)$ & 3 \\
\oo & ${1\leftarrow0}$ & $^OO$ & $(N, N)$ & 29 \\
\oo & ${1\leftarrow0}$ & $^OP$ & $(N, N)$& 17 \\
\oo & ${1\leftarrow0}$ & $^OP$ & $(N-1, N)$& 17 \\
\oo & ${2\leftarrow0}$ & $^QQ$ & $(N, N)$ & 19 \\
\oop & ${1\leftarrow0}$ & $^QQ$ & $(N, N)$ & 19 \\
\oop & ${1\leftarrow0}$ & $^SS$ & $(N, N)$ & 5 \\
\oop & ${1\leftarrow0}$ & $^OO$ & $(N, N)$ & 6 \\
\oopp & ${1\leftarrow0}$ & $^QQ$ & $(N, N)$ & 23 \\
\oopp & ${1\leftarrow0}$ & $^SS$ & $(N, N)$ & 19 \\
\oopp & ${1\leftarrow0}$ & $^OO$ & $(N, N)$ & 17\B \\
\hline\hline
\end{tabular}
\end{table}

\subsection{The 4LGSF lasing wavelengths}\label{sec:stability}

It was suggested that Raman lines can be used to assess the spectral accuracy of {\espr} \citep[see][]{Vogt2019a}. Doing so from the combination of 92 individual exposures acquired over 7 consecutive nights requires that the Raman signal be spectrally stable over this time span. The presence of wind (i.e., air molecules moving along the observing line-of-sight with respect to the rest frame of UT4) could lead to a varying spectral shift of the observed Raman lines between the different {\espr} exposures. A laser photon with frequency $\nu_{\lbeam}$ in the rest frame of UT4 will appear Doppler-shifted to a frequency $\nu^\prime$ to air molecules moving with a velocity $w_{\rm LoS}$ along the line-of-sight, where:
\begin{equation}
    \nu^\prime = \nu_{\lbeam}\left(1-\frac{w_{\rm LoS}}{c}\right)
\end{equation}
with $c=299792458$\,m\,s$^{-1}$ the speed of light, and $w_{\rm LoS}$ taken to be positive for molecules moving away from UT4. The frequency of this photon after being Raman-scattered is:
\begin{equation}
\nu^{\prime\prime} = \nu^\prime + \Delta\nu = \nu_{\lbeam}\left(1-\frac{w_{\rm LoS}}{c}\right)+\Delta\nu
\end{equation}
with $\Delta\nu$ the Raman-shift associated to the specific molecular transition (de-)excited through the inelastic scattering process. With the exception of the pure-rotational anti-Stokes lines, all the Raman lines observed in the {\espr} spectrum have $\Delta\nu<0$\,cm$^{-1}$. The Raman-scattered photon will then itself appear Doppler-shifted to a frequency $\nu_{\rm obs}$ in the UT4 rest frame, where:
\begin{equation}
    \nu_{\rm obs}=\nu^{\prime\prime}\frac{c}{c+w_{\rm LoS}}=\nu_{\lbeam}\frac{c-w_{\rm LoS}}{c+w_{\rm LoS}}+\Delta\nu\frac{c}{c+w_{\rm LoS}}
\end{equation}
We are here assuming that the laser photons are emitted along the optical axis of the UT4 telescope: a reasonable assumption given that the inclination of the laser beams for our observations is $\alpha\approx89.985^{\circ}$ (see Sec.~\ref{sec:foc-alt}).\\

When measuring the position of the Raman lines in the {\espr} spectrum, this Doppler-shift will result in a pseudo-redshift $zc$ of the lines compared to their expected frequency $\nu_{\rm ref}=\nu_{\lbeam}+\Delta\nu$, where:
\begin{eqnarray}
 zc &=& \frac{\lambda_{\rm obs}-\lambda_{\rm ref}}{\lambda_{\rm ref}} = \frac{\nu_{\rm ref}-\nu_{\rm obs}}{\nu_{\rm obs}}\nonumber\\
 &=& w_{\rm LoS}\frac{2\nu_{\lbeam}+\Delta\nu}{\nu_{\lbeam}\left(1-\frac{w_{\rm LoS}}{c}\right)+\Delta\nu}\nonumber\\
  &\approx& w_{\rm LoS}\left(2-\frac{\Delta\nu}{\nu_{\lbeam}+\Delta\nu}\right)\label{eq:zc}
\end{eqnarray}

The largest Raman shifts observed in the {\espr} spectra are associated to the $\nu_{2\leftarrow0}$ rotational-vibrational band of {\oo} (see Sec.~\ref{sec:deep}), which corresponds to $\Delta\nu\approx-3000$\,cm$^{-1}$. On the blueshifted side, pure-rotational Raman lines of the anti-Stokes branch can be identified only up to $\Delta\nu\approx+250$\,cm$^{-1}$. The \textit{Raman component} of $zc$ is thus small for all the lines in the {\espr} data, with:
\begin{equation}
    -0.02\lesssim\frac{-\Delta\nu}{\nu_{\lbeam}+\Delta\nu}\lesssim0.2
\end{equation}

The intensity and direction of the wind along the line-of-sight $w_{\rm LoS}$ can be estimated from the projection of the horizontal wind vector (derived from the ERA5 re-analysis dataset, see Sec.~\ref{sec:atmo-state})  at the altitude $H_{\rm obs}$ onto the UT4 pointing direction:
\begin{equation}
    w_{\rm LoS} = \vec{w}_{H_{\rm obs}}\cdot\vec{\zeta},
\end{equation}
where:
\begin{equation}
\vec{w}_{H_{\rm obs}}=\begin{pmatrix}w_u\\w_v\\0\end{pmatrix},\ \vec{\zeta}=\begin{pmatrix}\cos\psi\cos\phi\\\cos\psi\sin\phi \\ \sin\psi \end{pmatrix},
\end{equation}
with $\phi$ the UT4 azimuth measured counter-clockwise from the East, $\psi\in[0^{\circ}, 90^{\circ}]$ the UT4 pointing altitude, and $w_u$ ($w_v$) the ERA5 re-analysis eastward (northward) component of the wind at a geometric altitude $H_{obs}$. $w_u>0$ ($w_v>0$) indicates wind moving towards the East (North). With these conventions, the link between $\phi$ and the telescope azimuth provided in ESO FITS files is:
\begin{equation}
\phi=-(90+{\tt HIERARCH\ ESO\ TEL4\ AZ })
\end{equation}

\begin{figure}[htb!]
\centerline{\includegraphics[scale=0.5]{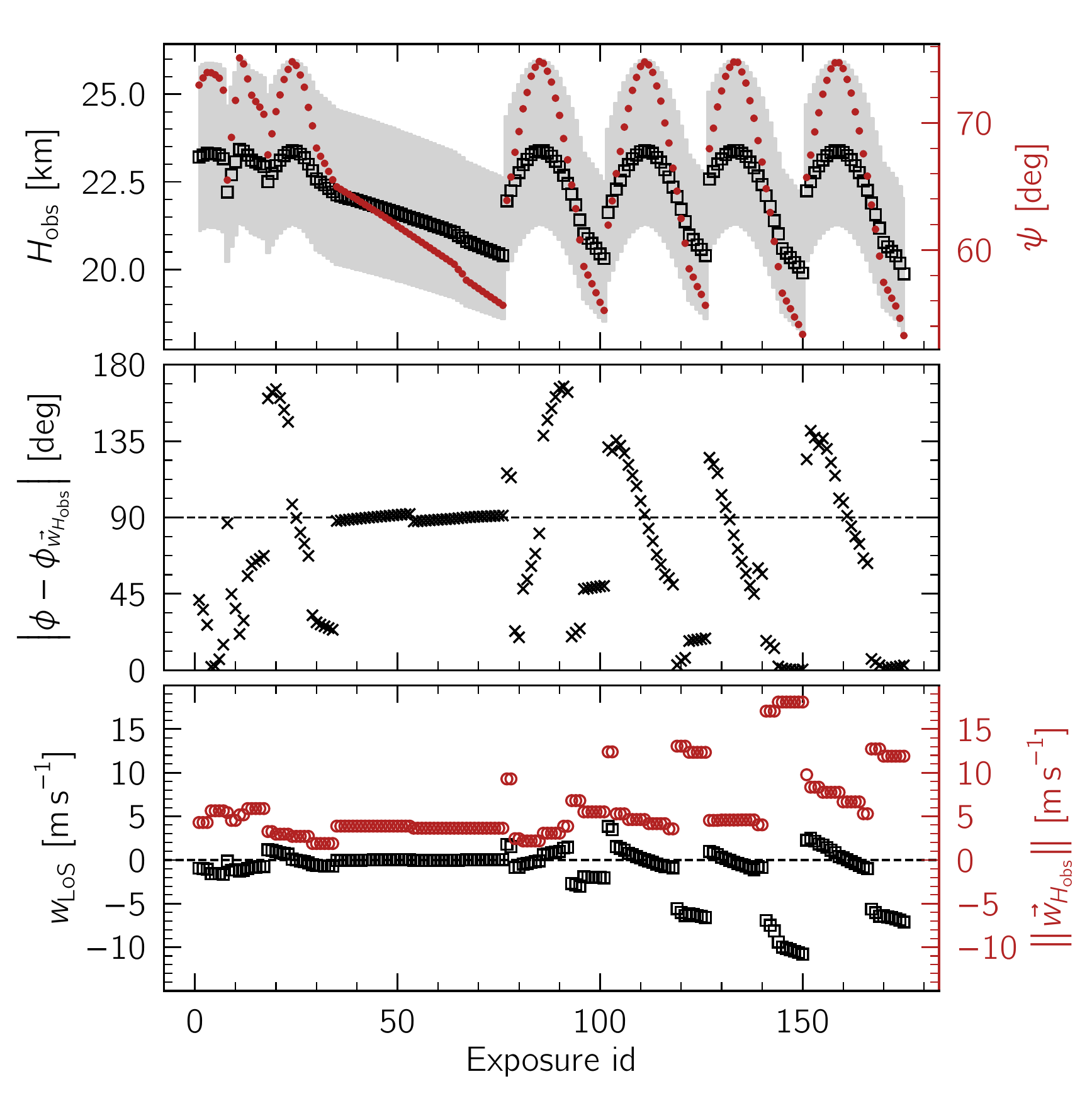}}
\caption{Top: geometric altitude $H_{\rm obs}$ associated to a line-of-sight focus distance $D=21.5$\,km for all the 175 {\espr} exposures acquired during the observing run. The shaded area shows the projected interval [$D-\Delta_D^-$; $D+\Delta_D^+$]. The telescope altitude $\psi$ is shown in red. Middle: angular offset between the UT4 pointing direction $\phi$ and the horizontal wind direction $\phi_{\vec{w}_{H_{\rm obs}}}$ at the altitude $H_{\rm obs}$, derived from the ERA5 re-analysis dataset. Bottom: wind speed along the UT4 line-of-sight $w_{\rm LoS}$ (black) associated to the total horizontal wind speed $\left\Vert\vec{w}_{H_{\rm obs}}\right\Vert$ at the geometric altitude $H_{\rm obs}$. }\label{fig:wlos}
\end{figure}

The value of $w_{\rm LoS}$ for every {\espr} exposure acquired during our observing run is presented in Fig.~\ref{fig:wlos}. With a vertical resolution of 2--3\,km at the altitudes of interest and a temporal resolution of 1\,hr, the ERA5 re-analysis dataset cannot be used to perform a detailed modelling of the wind shifts present in every {\espr} exposures, let alone correct them with a level of accuracy of $<10$\,m\,s$^{-1}$. It does nonetheless provide a reasonable estimate of the direction and order of magnitude of the redshifts $zc$ to be expected. This analysis indicates that the altitudes observed with {\espr} are located systematically above the jet stream layer, where horizontal wind speeds always remained below $\sim$15\,m\,s$^{-1}$ during our observing run. Accounting for the projection onto the telescope line-of-sight, the majority of our observations are thus unlikely to have been subjected to on-axis wind speeds larger than 5\,m\,s$^{-1}$. For the $\mathfrak{set\ 1}$ exposures specifically, 77 ($\equiv$84\%) exposures have $\left|w_{\rm LoS}\right|<2.5$\,m\,s$^{-1}$, 82 ($\equiv$89\%) exposures have $\left|w_{\rm LoS}\right|<5$\,m\,s$^{-1}$, and 92 ($\equiv$100\%) exposures have $\left|w_{\rm LoS}\right|<10$\,m\,s$^{-1}$.\\

\begin{figure*}[htb!]
\centerline{\includegraphics[scale=0.5]{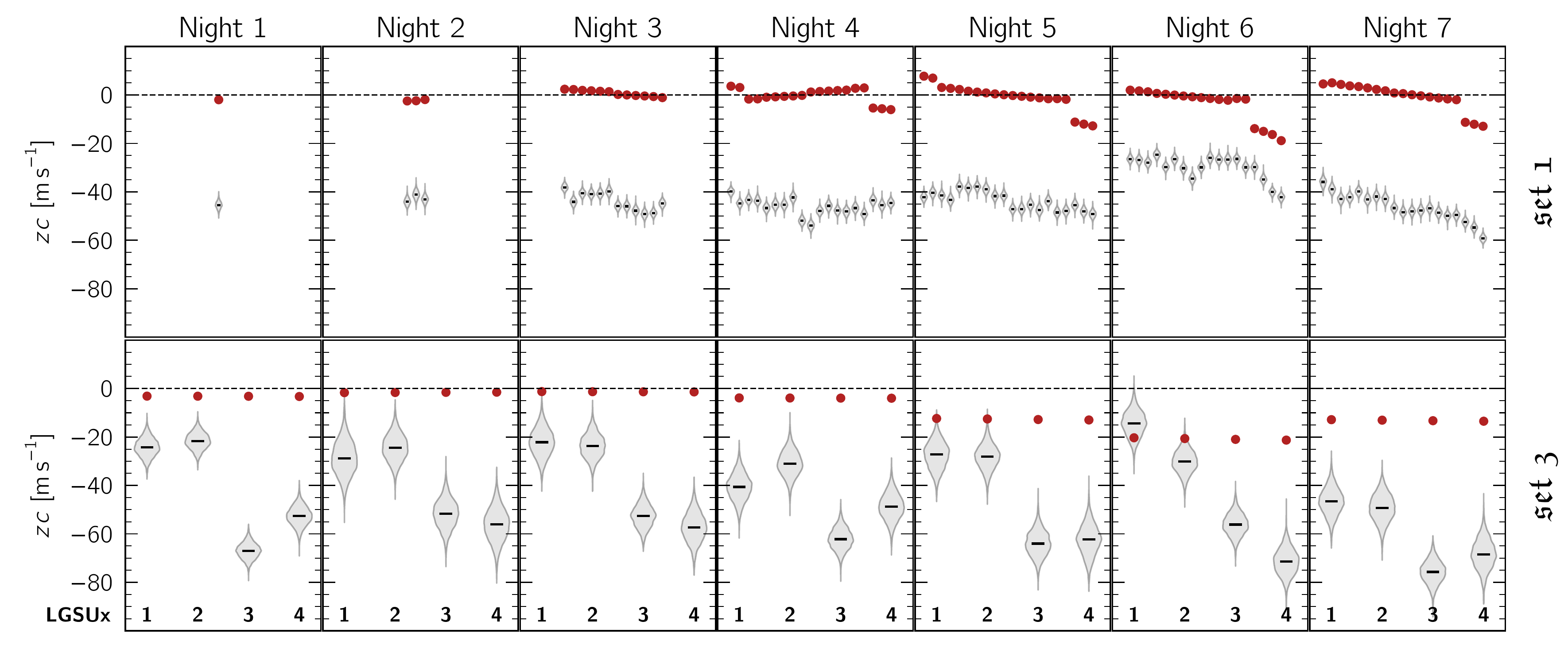}}
\caption{Measured redshift $zc$ of the pure-rotational Raman line associated to the {\oo} $^SS(19,19)$ transition, for every $\mathfrak{set\ 1}$ and $\mathfrak{set\ 3}$ exposure, split by observing night. The grey violin symbols trace our full sampling of the posterior distribution of $zc$ for this line in every exposure, with the median value marked using a black bar. Each measurement can be compared to the redshift that can be expected from the wind speed at the time and location of the observations following Eq.~\ref{eq:zc}, which is shown using red dots. For each $\mathfrak{set\ 3}$ exposure, the LGSU that was propagating (and thus observed) is indicated.}\label{fig:zc}
\end{figure*}

To measure the actual redshift $zc$ of Raman lines in the {\espr} spectra, we adopt a line model composed of 3 Gaussian components, where the outer two are located exactly $1.713$\,GHz on either-side of the central one, and have their peak intensity set to 1/9 of the central component. In other words, we assume the 4LGSF re-pumping side-lines are perfectly tied to the central line, both spectrally and in intensity. For each $\mathfrak{set\ 1}$ and $\mathfrak{set\ 3}$ individual exposure, we perform a Markov-Chain Monte-Carlo sampling of the individual posterior distribution of the parameters associated to a given line, using a least-square likelihood and flat priors with a lower bound of 0 for the line intensities and dispersion. For consistency with \cite{Vogt2019a}, we adopt $\lambda_{\lbeam}=5891.59120$\,{\AA} for the central lasing wavelength of all LGSUs. The results for the pure-rotational line associated to the {\oo} $^SS(19,19)$ transition (with a Raman shift of $\Delta\nu=-117.555334$\,cm$^{-1}$) are presented in Fig.~\ref{fig:zc}. For each exposure, a violin symbol is used to trace the full extent of our posterior distribution sampling of the line apparent velocity shift, together with the velocity shift expected from the wind alone (following Eq.~\ref{eq:zc}).\\

The slopes of the $\mathfrak{set\ 1}$ $zc$ measurements appear broadly consistent with the gradient expected from UT4 slowly turning towards the West (into the wind) as it tracks the target field across meridian. There is, however, a clear blueshift of $-40\pm5$\,m\,s$^{-1}$ in the line position, which is unexplained by the wind. This offset is constant for every observing night, albeit somewhat less strong for night 6. Most plausibly, this difference is a consequence of the reboot of the 4LGSF system (and lack of subsequent wavelength meter calibration) on the first half of that specific night.\\

Light from all four LGSUs is present in $\mathfrak{set\ 1}$ exposures, albeit in varying mixing ratios depending on the exposure (see Sec.~\ref{sec:acq}). The $\mathfrak{set\ 3}$ exposures, on the other hand, target a single LGSU each. They reveal that the blueshift $zc$ observed in $\mathfrak{set\ 1}$ exposures results in fact from the combination of different offsets for the different LGSUs: $-25\pm5$\,m\,s$^{-1}$ for LGSU1/2, and $-55\pm5$\,m\,s$^{-1}$ for LGSU3/4. In other words, the different LGSUs of the 4LGSF do not all lase at the same wavelength, and none of them lase at the nominal wavelength of $\lambda_{\lbeam}=5891.59120$\,\AA.\\

This explains the LGSU-dependant Raman line blueshifts detected consistently over 7 distinct nights. Alternative causes (meteorological/instrumental/technical/software) can all be ruled out given:
\begin{enumerate}
    \item the extreme spectral stability of {\espr} (by design),
    \item the time difference of no more than 3--4 minutes between consecutive $\mathfrak{set 3}$ exposures in a given quadruplet, with no difference in the instrumental/telescope setup other than a switch of the propagating LGSUs,
    \item the use of LFC calibration frames acquired within less than 2.5\,hr for all exposure on-sky, and
    \item the very stable atmospheric conditions over the entire duration of the observing run. 
\end{enumerate}

 We thus introduce $\epsilon_{\rm x}$ the difference between the LGSUs' true lasing frequency $\nu^{\mathsf{T}}_{\rm LGSUx}$ and $\nu_{\lbeam}=c\lambda_{\lbeam}^{-1}$:
 \begin{equation}
     \epsilon_{\rm x} = \nu^{\mathsf{T}}_{\rm LGSUx} - \nu_{\lbeam}
 \end{equation}
Similarly to the wind in Eq.~\ref{eq:zc}, $\epsilon_{\rm x}$ can be linked to the observed line redshift $zc$ via:
\begin{eqnarray}
zc&=&\frac{\nu_{\lbeam}-\nu_{\rm LGSUx}^{\mathsf{T}}}{\nu_{\rm LGSUx}^{\mathsf{T}}}\nonumber\\
  &=&\frac{-\epsilon_{\rm x}c}{\nu_{\lbeam}+\epsilon_{\rm x}+\Delta\nu},
\end{eqnarray}
such that:
\begin{eqnarray}
\epsilon_{\rm x} &=& -\frac{\nu_{\lbeam}+\Delta\nu}{1+\frac{c}{zc}}\nonumber\\
&\approx& -\left(\nu_{\lbeam}+\Delta\nu\right)\frac{zc}{c}
\end{eqnarray}

From our observations, we find that:
\begin{itemize}
    \item $\epsilon_{\rm 1}\approx\epsilon_{\rm 2}\approx+43\pm10$\,MHz $\cong-50\mp10$\,fm, and 
    \item $\epsilon_{\rm 3}\approx\epsilon_{\rm 4}\approx+94\pm10$\,MHz $\cong-109\mp10$\,fm.

\end{itemize}

The offset of $\sim$50\,MHz between LGSU1/2 and LGSU3/4 appears consistent with dedicated measurements acquired during the commissioning of the 4LGSF \citep[\url{https://www.eso.org/sci/libraries/SPIE2016/9909-27_slides.pdf};][]{Hackenberg2016}. The values of these spectral shifts are several times larger than the observed (relative) spectral stability of the LGUs \citep[better than $\pm3$\,MHz at the 1-$\sigma$ level over hours, see][]{Friedenauer2012}. The formal ESO requirements for the LGU's peak-to-peak laser emission wavelength variation over 14\,h is 80 \,MHz \citep[see][]{Enderlein2014}. The values of $\epsilon_{\rm 1}$, $\epsilon_{\rm 2}$, $\epsilon_{\rm 3}$, and $\epsilon_{\rm 4}$ also remain significantly smaller than the thermal broadening \citep[of the order of 1\,GHz;][] {Bradley1992,Holzlohner2010} of the atomic transitions of sodium atoms to be excited at an altitude of $\sim$90\,km to create laser guide-stars. They are thus of no consequences for adaptive optics purposes. They do, however, prevent any formal assessment of the spectral accuracy of {\espr} at the required level of 10\,m\,s$^{-1}$.


\section{Conclusions}\label{sec:conclusions}

We have used the high-resolution optical \'echelle spectrograph {\espr} to acquire, from the astronomical observatory of Cerro Paranal, a 15.3\,hr-deep spectrum of the four 4LGSF 589\,nm-sodium laser beams with a resolution of $\lambda/\Delta\lambda\approx140000\approx0.04$\,\AA~$\approx 0.12$\,cm$^{-1}$. This spectrum reveals the pure-rotational and rotational-vibrational Raman signature of the atmosphere at an altitude of 23\,km above sea level. It extends from 3775\,\AA\ to 7895\,\AA, equivalent to a Raman shift range of $+9517$\,cm$^{-1}$ to $-4307$\,cm$^{-1}$. It contains pure-rotational and rotational-vibrational Raman lines from the most abundant molecules in the atmosphere at this altitude, including their isotopes: \nn, \nnp, \oo, \oop, \oopp, and \coo. \\ 

The depth (i.e., the signal-to-noise) and spectral accuracy of these observations is such that professional observatories can treat the catalogue of identified Raman lines as exhaustive (over the specific Raman shift range, for the observed molecules) for the purpose of correcting Raman contamination of astronomical data. It must be noted than in addition to the 865 Raman transitions identified in the spectrum, a handful of emission line-like features (including 22 lines within $\pm56$\,\AA\ $\approx$ $\mp$160\,cm$^{-1}$ of the main laser line) remain unidentified at this stage. Some of them at least exhibit a \textit{Raman-like} distribution. However, the complex nature of the artefacts resulting from the strong saturation of the main laser line prevents us from ruling out an artificial origin without the formal identification of the molecular transitions. All these unidentified lines are at least 4.5 orders of magnitude fainter than the main laser lines, and should thus be of no concern to all but the most demanding astronomical observations.\\

The detection of spectral shifts between the different LGSUs of the 4LGSF prevent us from formally assessing the spectral accuracy of {\espr} as proposed by \citep[][]{Vogt2019a}. The main limitation lies in the unconstrained mix of light from the different LGSUs in the $\mathfrak{set\ 1}$ exposures, that cannot be disentangled prior to their combination. For the purpose of measuring the radial velocity accuracy of ultra-high resolution spectrographs like {\espr} at a level of 10\,m\,s$^{-1}$, independent knowledge of the lasing frequency with an accuracy $<10$\,MHz$\approx$10\,fm is paramount. The assessment of the accuracy of {\espr} by means of the Raman lines will thus remain out of reach until the lasing frequency of the different LGSUs can be measured down to that level. Our data indicate that the main lasing frequency of LGSU1/2 (LGSU3/4) is blueshifted by $+43\pm10$\,MHz\,$\cong-50\mp10$\,fm ($+94\pm10$\,MHz\,$\cong-109\mp10$\,fm) from $\lambda_{\lbeam}=5891.59120$\,\AA.\\

The fact that our observations reveal the presence of spectral shifts in the 4LGSF lasing frequencies at the level of a few tens of MHz nonetheless demonstrates how Raman lines can be exploited by professional observatories as wavelength references. Their observations is not straightforward. Short of the insertion of a notch filter in the optical path, the important contrast between the main laser line and the Raman lines will inevitably lead to the strong saturation of the former, for optical astronomical instruments with a broad spectral coverage. The resulting artefacts (see Sec.~\ref{sec:artefacts}) can pose real challenges for the subsequent analysis of the data. To reach the highest level of accuracy, the influence of the wind must also be taken carefully into account in the design of the observations. By observing at high elevations, outside from any jet stream, and at azimuths perpendicular to the dominant atmospheric currents, the spectral redshifts caused by wind can be kept within $\pm5$\,m\,s$^{-1}$ (see Sec.~\ref {sec:stability}).\\ 

Provided their lasing wavelength can be characterized with sufficient accuracy, the ever-increasing presence of laser guide-star systems at professional astronomical observatories world-wide could also prove interesting for atmospheric research. A high-resolution spectrograph like {\espr} can for example lead to a factor $\sim$10 improvement in the measured Raman transitions of \nn\ and \nnp\ (see Appendix~\ref{app:linelist}). The large collecting areas of ground-based astronomical telescopes could also enable the study and monitoring of very faint Raman signals, for example that of oxygen isotopes up to the highest layers of the atmosphere \citep[]{Wiesemeyer2023}.


\begin{acknowledgments}
\footnotesize
We are very grateful to A.~Cabral and J.~P.~Coelho for simulating the observational setup described in Sec.~\ref{sec:obs} using their \textsf{Zemax}$^{\copyright}$ model of UT4, its Coud\'e train, and the {\espr} instrument. We thank Telescope \& Instrument Operator Extraordinaire Diego Parraguez, Night Astronomer Anita Zanella, and Instrument Gurus Jos\'e-Luis Álvares, Juan Beltran Peña, and Álvaro Díaz for their infallible support during the observing run. We also acknowledge the courtesy of the {\muse} Consortium (with visitor time awarded immediately prior to our observations) for turning the telescope over to us on-time every night. We thank R.~Matthey de l'Endroit for enlightening discussions and valuable suggestions regarding this article, as well as the anonymous reviewer for his/her constructive feedback. This research has made use of the following Python packages:
\textsf{aplpy} \citep[an open-source plotting package for \textsf{Python};][]{Robitaille2012},
\textsf{astroplan} \citep[][]{Morris2018}
\textsf{astropy} \citep[a community-developed core \textsc{Python} package for Astronomy;][]{AstropyCollaboration2013,AstropyCollaboration2018},
\textsf{astroquery} \citep[a package hosted at \url{https://astroquery.readthedocs.io} which provides a set of tools for querying astronomical web forms and databases;][]{Ginsburg2017,Ginsburg2019},
\textsf{dfitspy} \citep{Thomas2019},
\textsf{emcee} \citep[][]{Foreman-Mackey2013},
\textsf{fcmaker} \citep[a \textsc{Python} module to create ESO-compliant finding charts for OBs on \textit{p2};][]{Vogt2018a,Vogt2018b},
\textsf{matplotlib} \citep{Hunter2007},
\textsf{numpy} \citep[][]{Harris2020},
\textsf{pandas} \citep[][]{McKinney2010, ThePandasDevelopmentTeam2021},
\textsf{scipy} \citep[][]{Virtanen2020},
\textsf{spectres} \citep[][]{Carnall2017}, and
\textsf{specutils} \citep{Earl2020}. This research has also made use of the \textsf{Aladin} interactive sky atlas \citep{Bonnarel2000}, of \textsf{SAOImage ds9} \citep[developed by Smithsonian Astrophysical Observatory][]{Joye2003}, and of NASA’s Astrophysics Data System. A portion of this work has been carried out within the framework of the National Centre of Competence in Research PlanetS supported by the Swiss National Science Foundation. PFi acknowledges the financial support of the SNSF. Other portions of this article present research carried out at the Jet Propulsion Laboratory, California Institute of Technology, under contract with the National Aeronautics and Space Administration. Government sponsorship is acknowledged by SYu. The article contains modified Copernicus Climate Change Service information 2022. Neither the European Commission nor ECMWF is responsible for any use that may be made of the Copernicus information or data it contains. Based on observations made with ESO Telescopes at the La Silla Paranal Observatory under P.Id.~4104.L-0074(A). All the observations described in this article are freely available online from the ESO Science Archive Facility. No detector was harmed during the astronomical observations presented in this article.
\end{acknowledgments}


\appendix

\section{Important acquisition steps for P.Id.~4104.L-0074}\label{sec:acq-seq}

The {\espr} observations of the 4LGSF up-link laser beams presented in this article were highly non-standard. For reproduceability purposes, we list here important steps in the acquisition and observing sequence.\\

\paragraph*{\bf Calibrations:}

\begin{enumerate}

    \item Perform manual wavemeter calibrations for all LGSUs every day of the run, in the afternoon.
    
    \item Execute a complete set of LFC and ThAr wavelength calibrations as close as possible before (after) the start (the end) of the observations. We systematically executed a set of \texttt{FP\_FP}, \texttt{THAR\_FP}, \texttt{FP\_THAR}, \texttt{THAR\_THAR}, \texttt{LFC\_FP}, and \texttt{FP\_LFC} calibrations.
\end{enumerate}

\paragraph*{\bf Acquisition steps:}
\begin{enumerate}

\item Start the telescope and {\espr} preset normally. Proceed until {\espr} asks to select the target.

\item Use the LPC to manually place all the lasers in the wide-field mode asterism, and apply the computed corrections. Then, move the lasers to the narrow-field mode asterism.

\item  The laser guide-stars (at $\sim$90\,km of altitude) are now visible out-of-focus as four donuts of $\sim$10\,arcsec in diameter in the {\espr} TCCD. Change the TCCD density filter to the minimum value. 

\item Stop the telescope Active Optics cycle after a few corrections have been applied.

\item Offset the focus of the telescope by moving the secondary mirror by a few mm to bring the up-link beams in focus. The focus nominal value was $-20$\,mm: we used an offset of $+10$\,mm to reach a final focus position of $-10$\,mm. The telescope guide-star gets lost in the process, and telescope guiding stops.

\item Send absolute Alt-Az presets to the 4LGSF to move the laser beams into position. Monitor the changes using the {\espr} TCCD. For reference, here is a reasonable guess for the offsets required to cross the beams over the {\espr} fiber A:
\begin{itemize}
 \item LGSU1: Alt: $+35$\,arcsec; Az: $-13$\,arcsec
 \item LGSU2: Alt: $-40$\,arcsec; Az: $-13$\,arcsec
 \item LGSU3: Alt: $-40$\,arcsec; Az: $+02$\,arcsec
 \item LGSU4: Alt: $+33$\,arcsec; Az: $+05$\,arcsec 
\end{itemize}

Loosing the beams while moving them is extremely likely, when applying large positional shifts or focus changes. Applying a series of small successive focus changes and positional offsets is thus preferable, at least for the first acquisition.

\item  With the four up-link laser beams crossed over the {\espr} fiber,  continue with the {\espr} acquisition. Disable the field stabilization when asked.

\item Execute the laser observations. Monitor the motions of the laser beams with respect to the fiber using the {\espr} TCCD. Use the {\espr} exposure meter to monitor the flux received in the fiber A.

\item Pause the observing every 20 minutes. Sequentially toggle the propagation of individual laser guide-stars on/off. For each, adjust the pointing as necessary by sending manual offsets to the LGSU.  

\item Every hour, pause the observing, revert the telescope focus to the sky ($\equiv$ infinity), and apply a few cycles of active optics corrections. This helps account for the gravity vector changing over time for the telescope.

\end{enumerate}

\newpage
\section{Observing log}\label{app:obslog}

We list in Table~\ref{tbl:obs-log} the complete list of raw {\espr} exposures acquired under P.Id.~4104.L-0074. All these files are freely accessible from the ESO Science Archive Facility. The status of the different LGSUs (either propagating, or not) is specified explicitly for each exposure. This table is accessible electronically as a VizieR \footnote{https://doi.org/10.26093/cds/vizier} catalogue \citep{Ochsenbein2000} at the Centre de Données astronomiques de Strasbourg (CDS) via anonymous ftp \footnote{cdsarc.u-strasbg.fr (130.79.128.5)} or over https \footnote{https://cdsarc.unistra.fr/viz-bin/cat/Jother/PhRvR}.\\

{\LTcapwidth=\textwidth
\begin{longtable*}{>{\footnotesize}r >{\footnotesize}c >{\footnotesize}c >{\footnotesize}c >{\footnotesize}c >{\footnotesize\centering}m{4mm} >{\footnotesize\centering}m{4mm} >{\footnotesize\centering}m{4mm} >{\footnotesize\centering}m{4mm} >{\footnotesize}c >{\footnotesize}l}
\caption{Observing log for ESO P.~Id.~4104.L-0074. For each exposure, the status of each LGSU is specified explicitly: either \textit{propagating} (\faToggleOn), or \textit{not propagating} (\faToggleOff). The $\mathfrak{set}$ value identifies different groups of exposures discussed in the article.}\label{tbl:obs-log}\\
\midrule\midrule
\# & ESO archive filename & UT4 altitude & Focus & Exp. time & \multicolumn{4}{c}{LGSUx status} & $\mathfrak{set}$ & Comments\\
& & [$^{\circ}$] & [mm] & [s] & 1 & 2 & 3 & 4 & &\\ 
\midrule
\endfirsthead
\caption{cont.}\\
\midrule
\midrule
\# & ESO archive filename & Altitude & Focus value & Exp. time & \multicolumn{4}{c}{LGSUx status} & $\mathfrak{set}$ & Comments\\
& & [$^{\circ}$] & [mm] & [s] & 1 & 2 & 3 & 4 & &\\ 
\midrule
\endhead
\midrule
\endfoot
\midrule\midrule
\endlastfoot
\multicolumn{11}{c}{\leaders\hbox{\rule[0.3em]{.1pt}{0.4pt}}\hfill\mbox{} Night 1 \leaders\hbox{\rule[0.3em]{.1pt}{0.4pt}}\hfill\mbox{}}\\
1 & ESPRE.2019-10-22T05:02:07.619.fits & $72.966$ & $-10.0$ & $300.0$ & \faToggleOn & \faToggleOn & \faToggleOn & \faToggleOn & $\mathfrak{-}$ &\\ 
2 & ESPRE.2019-10-22T05:11:23.619.fits & $73.528$ & $-10.0$ & $600.0$ & \faToggleOn & \faToggleOn & \faToggleOn & \faToggleOn & $\mathfrak{1}$ &\\ 
3 & ESPRE.2019-10-22T05:24:43.346.fits & $73.955$ & $-10.0$ & $600.0$ & \faToggleOff & \faToggleOff & \faToggleOff & \faToggleOff & $\mathfrak{-}$ &\\ 
4 & ESPRE.2019-10-22T05:36:16.411.fits & $73.937$ & $-10.0$ & $300.0$ & \faToggleOn & \faToggleOff & \faToggleOff& \faToggleOff & $\mathfrak{3}$ &\\ 
5 & ESPRE.2019-10-22T05:42:36.409.fits & $73.774$ & $-10.0$ & $300.0$ & \faToggleOff & \faToggleOn & \faToggleOff & \faToggleOff & $\mathfrak{3}$ &\\ 
6 & ESPRE.2019-10-22T05:48:51.138.fits & $73.508$ & $-10.0$ & $300.0$ & \faToggleOff & \faToggleOff & \faToggleOn & \faToggleOff & $\mathfrak{3}$ & Moonrise: 5:57 UT\\ 
7 & ESPRE.2019-10-22T06:02:57.681.fits & $72.560$ & $-10.0$ & $300.0$ & \faToggleOff & \faToggleOff & \faToggleOff & \faToggleOn & $\mathfrak{3}$ &\\
\end{longtable*}}
\newpage $ $\newpage $ $\newpage


\section{Raman lines identified in the 15.3\,hr-deep {\espr} spectrum of the 4LGSF laser beams}\label{app:linelist}

Table~\ref{tbl:lines} provides an exhaustive list of all the pure-rotational and rotational-vibrational Raman lines that could be identified in the {\espr} spectrum described in Sec.~\ref{sec:deep}. This table is accessible electronically as a VizieR \footnote{https://doi.org/10.26093/cds/vizier} catalogue \citep{Ochsenbein2000} at the Centre de Données astronomiques de Strasbourg (CDS) via anonymous ftp \footnote{cdsarc.u-strasbg.fr (130.79.128.5)} or over https \footnote{https://cdsarc.unistra.fr/viz-bin/cat/Jother/PhRvR}. The lines were not identified individually, but rather by sets (see Table~\ref{tbl:rot-o2} and \ref{tbl:rotvib-o2} for the sets of O$_2$). We limited the lines included in a given set to quantum numbers for which the line has a signal-to-noise greater than 1. The only exceptions are the \oo\ rotational-vibrational lines in the sets $^SS\ (N,N-1)$ and $^OO\ (N, N-1)$. They cannot be identified beyond $N=1$ and $N=3$ (respectively) because of their blending with lines in the sets $^SS\ (N,N)$ and $^OO\ (N,N)$.\\

Strictly speaking, not every line listed in Table~\ref{tbl:lines} is \textit{visible} in the {\espr} spectrum presented in Fig.~\ref{fig:full}. Some lines are blended with stronger ones. Others are affected by saturation-induced artefacts. Several are also located in the immediate vicinity of the main laser line at $\Delta\nu=0$: a spectral region (reconstructed from the $\mathfrak{set\ 4}$ exposures) that has a smaller signal-to-noise. These lines are included in Table~\ref{tbl:lines} nonetheless, on the basis that 1) they belong to clearly identified theoretical sets found to match the data without evident deviation, and 2) they are surrounded by other lines in the same set that are visible in the spectrum.\\

The lines are listed in descending order of their Raman shift $\Delta\nu$, with the convention that $\Delta\nu=\nu_{\rm obs}-\nu_{\lbeam}$ with $\nu_{\rm obs}$ and $\nu_{\lbeam}$ the Raman-scattered and original laser photon frequencies. The corresponding vacuum wavelengths $\lambda_{\rm vac}$ are computed with respect to $\lambda_{\lbeam}=5891.59120$\,\AA\ via:
\begin{equation}\label{eq:shift_to_lam}
\lambda_{\rm vac} = c\left(\frac{c}{\lambda_{\lbeam}}-\Delta\nu_{\rm Hz}\right)^{-1},
\end{equation}
with $\Delta\nu_{\rm Hz}$ the Raman shift $\Delta\nu$ expressed in Hz. These wavelengths are quoted to 3 decimals only, in view of the uncertainty regarding the exact lasing frequencies of the 4LGSF LGUs (see Sec.~\ref{sec:stability}). Readers requiring accuracy down to the fm-level are urged to re-compute the values of $\lambda_{\rm vac}$ via Eq.~\ref{eq:shift_to_lam} using the best possible estimate of $\lambda_{\lbeam}$ available to them at the time.\\

It must be stressed that the list of Raman shifts provided in Table~\ref{tbl:lines} is not the result of a fitting procedure using the {\espr} data described in this article. These are all obtained from the literature, and merely compared against the {\espr} spectrum to identify the different emission lines in it. The shifts therefore have varying degrees of accuracy, with those associated to the O$_2$ molecule being the most accurate (and thus the most suitable for acting as wavelength references). The readers are referred to the original sources for specific details:
\begin{itemize}
\item {\bf \oo, \oop, \oopp:}\\
The Raman shifts for these molecules were obtained from a complex Hamiltonian model \citep[][]{Yu2012,Drouin2012,Drouin2013,Yu2014} that was used to simultaneously fit the microwave, THz, infrared, visible and ultraviolet
transitions of all six oxygen isotopologues. The model accounts for the $^3\Sigma$ electronic ground state of these molecules, which leads to a fine-structure splitting of the lines. The specific fine-structure level of each O$_2$ line in Table~\ref{tbl:lines} is specified as $^{\Delta N}\Delta J\ (J_{\rm lower},N_{\rm lower})$ alongside the vibrational levels. With an accuracy $<0.0001$\,cm$^{-1}$, these Raman shifts are the most accurate in Table~\ref{tbl:lines}.

\item {\bf \nn, \nnp:}\\
We treat these molecules as diatomic non-rigid rotators, and account for the vibrational stretching of the molecular bond driven by rotation to derive the Raman shifts for the different branches and values of the rotational quantum number $J=N$. The corresponding equations are laid out in details in the Appendix of \cite{Vogt2019a}, to which we refer the reader for further details. Unlike \cite{Vogt2019a}, however, we rely here on the molecular parameters from \cite{LeRoy2006}. The specific transition of each N$_2$ line in Table~\ref{tbl:lines} is identified in the form $\Delta J\ (J_{\rm lower})$. Their accuracy is $\sim0.001$\,cm$^{-1}$. One should note that in the case of anti-stoke lines, $J_{\rm lower}\not\equiv J_{\rm final}$, since the laser photons gain energy from excited molecules in those transitions. With our adopted notation, the first stokes and anti-stokes lines of a given molecule are thus both labelled as $S (0)$ \citep[for more details, see Sec.~8.2 in][]{Bernath2016}.

\item {\bf\coo:}\\
For the pure-rotational lines, we treat these molecules using the same approach as N$_2$, with the molecular constants from \citep[][]{Barrett1970}. The rotational-vibrational shifts, on the other hand, are taken from \cite{Tejeda1995}. With an accuracy $\sim0.1$\,cm$^{-1}$, these Raman shifts are the least accurate in Table~\ref{tbl:lines}.

\end{itemize}

The intensity of the Raman lines in a dedicated/contaminated observation acquired using an astronomical spectrograph will be highly dependent on:
\begin{enumerate}
    \item the technical specifications of the instrument, telescope, and laser system,
    \item the parameters of the laser \textit{collision}, and (to a much lesser extent),
    \item the meteorological conditions.
\end{enumerate}
In Table~\ref{tbl:lines}, we thus provide a measure of the order of magnitude difference between the intensity of the main laser line $I_{\lbeam}$ (visible through Rayleigh and Mie scattering at $\Delta\nu=0$) and the intensity of each Raman line $I$, in lieu of direct fluxes. This measure, computed as $\log\left(I_{\lbeam}/I\right)$, is directly estimated from the {\espr} spectrum presented in Fig.~\ref{fig:full}. It indicates, in relative terms, which Raman lines are the strongest, and thus most easily detectable. It must be stressed that this measure is indicative only, in that it does not rely on any complex line fitting scheme. No special treatment is applied to blended lines, the relative intensity of which will be overestimated. Overall, we identify Raman lines with intensities spanning 5 orders of magnitude.\\

{\LTcapwidth=\columnwidth
\begin{longtable}{>{\footnotesize}c >{\footnotesize}c >{\footnotesize}c >{\footnotesize}c >{\footnotesize}c >{\footnotesize}c}
\caption{Exhaustive list of all the Raman lines identified in the 15.3\,hr-deep {\espr} spectrum of the 4LGSF laser beams. The line wavelengths are quoted in vacuum. They are computed with respect to $\lambda_{\lbeam}=5891.59120$\,\AA. The value $\log(I_{\lbeam}/I)$ provides a crude measure of the order of magnitude difference between a given line and the main laser line visible through Rayleigh and Mie scattering at $\Delta\nu=0$. The larger this value, the fainter the Raman line. }\label{tbl:lines}\\
\midrule\midrule
 $\Delta\nu$ & $\lambda_{\rm vac}$ & $\log\left(\frac{I_{\lbeam}}{I}\right)$ & Molecule & \multicolumn{2}{c}{Transition} \\
 {[cm$^{-1}$]} & [\AA] &  &  &  & \\
\midrule
\endfirsthead
\caption{cont.}\\
\midrule
\midrule
 $\Delta\nu$ & $\lambda_{\rm vac}$ & $\log\left(\frac{I_{\lbeam}}{I}\right)$ & Molecule & \multicolumn{2}{c}{Transition} \\
 {[cm$^{-1}$]} & [\AA] &  &  &  & \\
\midrule
\endhead
\midrule
\endfoot
\midrule\midrule
\endlastfoot
+249.248396 & 5806.327 & 6.4 & \nn & $\nu_{0\leftarrow 0}$ & $S$ (30) \\ 
+241.431361 & 5808.964 & 6.4 & \oo & $\nu_{0\leftarrow 0}$ & $^{S}S\ (41,41)$ \\ 
+241.422811 & 5808.967 & 6.4 & \nn & $\nu_{0\leftarrow 0}$ & $S$ (29) \\ 
+233.588798 & 5811.611 & 6.2 & \nn & $\nu_{0\leftarrow 0}$ & $S$ (28) \\ 
+230.330204 & 5812.712 & 6.5 & \oo & $\nu_{0\leftarrow 0}$ & $^{S}S\ (39,39)$ \\ 
+225.746637 & 5814.261 & 6.2 & \nn & $\nu_{0\leftarrow 0}$ & $S$ (27) \\ 
+219.191412 & 5816.478 & 6.4 & \oo & $\nu_{0\leftarrow 0}$ & $^{S}S\ (37,37)$ \\ 
+217.896601 & 5816.916 & 5.9 & \nn & $\nu_{0\leftarrow 0}$ & $S$ (26) \\
\end{longtable}
}

\bibliographystyle{apsrev4-1_lim10}

\bibliography{bibliography_fixed}

\end{document}